\documentclass[aps,showpacs]{revtex4}
\usepackage{graphicx}% Include figure files
 \usepackage{amsmath,amssymb,pstricks,pst-plot,graphics}
\usepackage{psfrag}           
\usepackage{bm}% bold math
 \def\comment#1{}
   
 \newcommand{\nts}[1]{\tmspace{-}{#1\thinmuskip}{#1\txtmu}}
\begin{document}
\title{Microwave photoconductivity of a periodically modulated two-dimensional 
electron system: Striped states and overlapping Landau levels}
\author{J\"urgen  Dietel}
\affiliation{
Institut f\"ur Theoretische Physik, \\
 Freie Universit\"at Berlin, \\
 Arnimallee 14, D 14195 Berlin, Germany}

\begin{abstract}
We study the {\it dc}
  response of the striped state under microwave irradiation.
The striped state can be modeled by a two-dimensional electron system under the influence of a unidirectional periodic modulation potential, which was examined in  [J.~Dietel {\it et al.},
  Phys.~Rev.~ B {\bf 71}, 045329 (2005)]. 
We further extend our study of the periodically modulated system to the case of strongly overlapping Landau levels and calculate the dark  
 and  photoconductivities. 
The strength of the modulation potential serves as an additional parameter which allows to vary the overlap of the Landau levels independently of the filling fraction.
\end{abstract}

\maketitle

\section{Introduction}
Recent experiments \cite{Mani,Zudov,Yang,Dorozhkin,Willet1} on a
two-dimensional electron gas (2DEG) in weak magnetic fields show that these
systems have an almost vanishing longitudinal resistivity when the
microwave frequency $ \omega $ is (up to an additive constant)
approximately a multiple of the cyclotron frequency.  This triggered a
number of theoretical and experimental papers 
\cite{Dietel1,Mani,Zudov,Yang,Dorozhkin,Willet1,Andreev,
  Bergeret,Ryzhii,Durst, Shi, Lei, Vavilov1, Dmitriev1, Dmitriev2, Inarrea1, 
Auerbach1, Gumbs1,   Robinson1, Joas1, Mani2, Dorozhkin2, Ryzhii2, Lei2}.  As is shown in
Ref.\ \onlinecite{Andreev}, zero resistance  
states can be due to a negative microscopic diagonal
conductivity. There are two mechanisms producing  negative microscopic
longitudinal conductivity: The first mechanism, known as 
{\em displacement mechanism}, relies on
disorder-assisted absorption and emission of microwaves \cite{Durst,
  Shi, Lei, Vavilov1, Ryzhii}.  Depending on the detuning $ \Delta
\omega = \omega_c - \omega $, where $\omega_c$ is the cyclotron
frequency, the displacement of the electrons is preferentially in or
against the direction of the applied {\it dc } field.
  An alternative mechanism, known as {\em distribution function mechanism}, 
relies on
the fact that microwave absorption leads to a change in the electronic
distribution function, which can result in a negative
photoconductivity \cite{Dorozhkin, Dmitriev1, Dmitriev2}. 
Detailed calculations
within the self-consistent Born approximation suggest that the latter
mechanism is larger by a factor $ \tau_{\rm in}/\tau_{\rm s}^* $ where
$ \tau_{\rm in} $ is the inelastic relaxation time and $ \tau_{\rm
  s}^* $ the single particle scattering time in the presence of the
magnetic field.

As is well known, the influence of a unidirectional periodic 
potential on a 2DEG 
in a homogeneous magnetic field leads to transport anisotropies \cite{Willet2}
as well as commensurability oscillations in the 
resistivity known as Weiss oscillations \cite{Weiss}. In such systems
the Landau level broadening is due to the periodic modulation 
induced by the additional potential.    
In a foregoing paper \cite{Dietel1}, we studied the dark conductivities 
as well 
as the microwave-induced photoconductivities for these systems 
using a Fermi's golden 
rule approach which goes back to Titeica in Ref. \onlinecite{Titeica1}.
The {\it dc} response of the microwave-driven system in the presence of 
a modulation potential 
is similar to the magnetoresistivity response
of a 2DEG to surface acoustic waves \cite{Robinson1}.

It is clear that one can calculate the current resulting from a
perturbation of the electron system by calculating the transition
rates.  It is easy to establish a current formula from the transition
rates between eigenstates when knowing their position expectation values
 \cite{Dietel1}.  Within the lowest order approximation,
we can calculate these transition rates by the help of Fermi's golden
rule.  Without microwave field, one can calculate the dark conductivity 
using the impurity 
potential as a perturbation. Under microwave irradiation, the 
microwave operator acts
as an additional perturbation.

In Ref. \onlinecite{Dietel1}, we carried out a calculation of 
the dark and the 
photoconductivity to the most obvious case 
where the periodic modulation wave length $a$ of the unidirectional 
field is small compared to the cyclotron radius $R_c$ and 
additional small to the inverse 
of the correlation length $ \xi $ times the square of the magnetic length, 
i.e. $ \ell_B^2/\xi $. Furthermore we restricted us to   
temperatures $ T $ large with respect to the Landau level 
width, $ 2 V_N$. $ N $ is the index of the 
topmost filled Landau level. Furthermore, we limited our calculation to the 
case that the amplitude of the unidirectional modulation potential 
$ \tilde{V} $ 
is smaller than $\omega_c$ such that 
we were able to restrict ourselves to first order perturbation theory in 
$ \tilde{V} $. Finally, we discussed explicitly 
the conductivities for the 
case $ \omega \approx \omega_c $ which is known to be the dominant regime 
for zero resistance states. 
We obtained in Ref. \onlinecite{Dietel1} that the 
photocurrent parallel to 
the direction of the wavevector of the unidirectional field is governed by the 
distribution function mechanism, which is larger than the 
contribution of the displacement mechanism by 
a factor $ \tau_{\rm in}/\tau_{\rm s}^* $. 
In the perpendicular direction, we find a strong enhancement of the 
displacement mechanism photocurrent such that in this case both 
contributions are of the same order.
In a recent paper \cite{Joas1}, we calculated 
the various conductivities in the regime 
$ \omega/\omega_c \ll 1$ which, experimentally, for the case 
without periodic modulation potential, has shown a strong suppression 
of the Shubnikov de-Haas oscillations \cite{Mani2, Dorozhkin2}, 
in accordance with our calculation. Further theoretical work on 
this subject was done in \cite{Ryzhii2,Lei2}.

There are other interesting applications of 
our formalism.
The above system can be used to model different types of 
anisotropic quantum Hall systems, e.g. the striped 
state system \cite{Moessner, Fogler}, which shows a 
more complex unidirectional modulation 
even without the application of a modulation potential. In the striped state, 
the modulation has a 
wavelength $ a \approx R_c $.  
The available transport experiments in this regime 
were carried out at temperatures much smaller than the 
Landau level spacing \cite{Lilly,Du2}, but without microwave irradiation.
We argue in this paper that interesting effects, among which 
 zero resistance states, can arise when the striped 
system is irradiated by microwaves. 
Motivated by the physics of the striped state,
we first calculate the various conductivities 
for the case  $ T \ll V_N $ and $ a  \ll \ell_B^2/\xi,R_c $.  
We obtain a strong filling fraction dependence of these 
conductivities, except for the displacement 
mechanism contribution in the perpendicular direction. 
In addition, we discuss 
the dark conductivity and the photoconductivities in the striped state regime 
$ a \approx R_c $ and obtain agreement of the {\it dc} (dark) 
conductivity in the striped state with prior calculations. 
  
In our calculations in Ref. \onlinecite{Dietel1} we assumed 
that the impurity contribution to the 
Landau level band width is negligible 
in comparison to the contribution from the
modulation potential. This is in contrast
to the experimentally realized system without unidirectional 
periodic modulation where the band width has its origin in 
impurity scattering.  
The band structure in this system depends mainly on the strength of the 
magnetic field. For weak magnetic fields, the system 
has highly overlapping Landau 
levels, while for strong  magnetic fields, the Landau levels are well   
separated. By a variation of the modulation potential amplitude $ \tilde{V} $ 
 -- which is possible in the usual Weiss oscillation experiments by 
tuning  the grid potential \cite{Weiss} --  it is possible to reach the  
overlapping Landau level regime without changing the magnetic field
and thus the effective filling fraction.  
It is physically plausible that the conductivity expressions 
parallel to the modulation potential 
should be in accordance to the system without unidirectional modulation
given e.g.  in Ref. \onlinecite{Vavilov1, Dmitriev1} up to numbers.  
This was shown explicitly in 
Ref. \onlinecite{Dietel1} for the non-overlapping Landau level regime,
i.e. $ V_N \ll \omega_c$. In this paper we show the same for the 
highly overlapping Landau level regime, i.e. $ V_N \gg \omega_c$.
We obtain also an oscillating photoconductivity but the explicit 
frequency dependence differ from the ones without modulation 
potential.  

For calculating the various conductivities in the case 
$ V_N \gg \omega_c $ we have to discuss the single particle 
eigenvalue problem of an electron in a
unidirectional periodic potential and a homogeneous magnetic field. 
This was done in Ref. \onlinecite{Dietel1} by solving 
the eigenvalue problem in first order perturbation theory in 
the oscillating potential strength, 
which is appropriate for $ V_N \ll \omega_c $. 
In this paper, we solve the eigenvalue problem 
without any restrictions to the relation between $ V_N $ and $ \omega_c  $ by 
discussing the WKB approximation of this eigenvalue problem. 
These results are used in Section VI to calculate the
conductivities for the case $ V_N \gg  \omega_c $.

We chose to provide a self-contained summary of all main results 
and a discussion of their implications relevant to experiments in 
the next section of this paper. 
 
This paper is organized as follows: 
Section II contains a summary of our results.  In section III we calculate 
the various  dark and photoconductivities for the distribution 
function and the displacement mechanism in the case where 
$ T \ll V_N \ll \omega_c $ and $ a  \ll \ell_B^2 /\xi,R_c $. In section IV 
we generalize our discussion to arbitrary $ a $ and $ R_c $, 
which is relevant for 
the discussion of the dark and microwave photoconductivity in the 
striped state. In section V, we discuss the WKB approximation 
of an electron in a homogeneous magnetic field and an additional periodic 
unidirectional potential. By using the results of section V, we discuss 
in section VI the dark as well as the photoconductivites for the case 
$ V_N \gg \omega_c $. The derivations of sections III-VI 
were carried out in the frequency regime $ \omega \approx \omega_c $ 
and for a linearly polarized microwave field in $x$-direction. In section VII,
we generalize these result
to an arbitrary microwave field of frequency $ \omega \gtrsim \omega_c $.   
We summarize our results in section VIII. 
Some technical details are deferred to a number of 
appendices.
In the remainder of this paper we use the convention 
$ \hbar =1 $. 

\section{Overview of the Results}

For clarity we point out once more that the explicit
conductivity formulas of sections III-VI are only valid 
for $ \omega \approx \omega_c $ and $ x $ polarization. 
We assume that the Landau level bandwidth due to impurity scattering is much 
smaller than the bandwidth due to the modulation potential. 
All conductivity formulas in this paper are expressed 
via the bandwidth $ 2 V_n $ of the $n$th Landau band (\ref{15}). 
The density of the states of the $ n$th Landau level 
$ \nu^*_n(\epsilon) $ and of the whole system $ \nu^*(\epsilon) $  
for $ V_N \ll \omega_c $ are given in (\ref{20})-(\ref{45}).  
The corresonding density of states $ \nu_{\Sigma}^*(\epsilon) $ 
in the large overlapping Landau level 
regime $ V_N \gg \omega_c $ is given in (\ref{2530})-(\ref{2535}). 
We have expressed the various conductivities by the help of the 
diffusion constants $ D_{xx} $, $ D_{yy} $ (\ref{103}) 
for $ V_N \ll \omega_c $ and $ D^{\Sigma}_{xx} $, 
$ D^{\Sigma}_{yy} $ (\ref{2565}) for  $ V_N \gg \omega_c $.
The single particle $ \tau^*_{\rm s}(\epsilon) $ 
and the transort time  $ \tau^*_{\rm tr}(\epsilon) $ is defined below 
(\ref{103}).  

As mentioned above, in our  paper Ref.\onlinecite{Dietel1}  
we carried out the calculation of the various 
dark conductivities, i.e. $ \sigma_{xx} $ parallel to the modulation 
and $\sigma_{yy}$ perpendicular to the modulation, as well as the 
photoconductivities $ \sigma^{\rm DP}_{xx} $, $ \sigma^{\rm DP}_{yy} $ 
of the displacement mechanism and 
$ \sigma^{\rm DF}_{xx} $, $ \sigma^{\rm DF}_{yy} $ of the distribution
function mechanism in the parameter range $ V_N \ll  T, \omega_c  \ll 
N \omega_c  $  and  $ a  \ll \ell_B^2 /\xi,R_c $. 
Under this condition, an electron can jump over  
many periods of the modulation potential by microwave-assisted 
impurity scattering.    
On the way to calculate the various conductivities for the striped state 
we extend in section III the derivations to temperature 
$ T \ll V_N \ll \omega_c $ but still $ a  \ll \ell_B^2 /\xi,R_c $. 
In this case the  dark and photoconductivites 
are  strongly filling 
 fraction dependent except for the photoconductivity of the 
displacement mechanism in the direction perpendicular 
to the periodic potential, i.e. for  $ \sigma^{\rm DP}_{yy} $. In the case 
of integer filling fraction, we obtain the same photoconductivities 
as calculated earlier for $ T \gg V_N $ 
in Ref. \onlinecite{Dietel1}. The expressions for the dark conductivities 
are given in (\ref{105}) and (\ref{110}). The results for the 
photoconductivites are given in (\ref{140}), (\ref{160}), (\ref{178})
and (\ref{190}). The photoconductivity expressions depend via 
the  functions $ A_1 $, $A_2 $, $ B_1 $, $ B_2 $ on the 
frequency  $ \Delta \omega =\omega_c -\omega $ and the filling fraction 
$ \Delta \mu=\mu-(N+1/2)\omega_c $. These functions 
are shown in Fig.~1 in the case of half integer upper Landau level 
filling $ \Delta \mu=0 $ most relevant 
for the striped state system. Corresponding functions for the integer 
filling system can be found in Ref. \onlinecite{Dietel1}.  
In (\ref{200}) we provide scaling relations for the 
photoconductivities of the displacement and distribution function mechanism 
in both directions.

In section IV we calculate the various conductivities without restrictions on 
$ a $ and $ R_c $. This is relevant for the striped state where 
$ a \approx R_c $. For impurity correlation length $ a \gg \ell_B^2/\xi $ 
or $ a \gg R_c  $, respectively,  
which is the strong forward scattering case in the striped state system,  
we obtain  (\ref{523}) and (\ref{526}) with (\ref{105}) for 
the dark conductivities in both directions.
The calculated dark conductivities fulfill the product rule  
(\ref{528}) which is in accordance 
with the well known semicircle law for half filled systems  
derived before for striped states.  
In the case of the photoconductivities one has to distinguish between the 
regimes where the absolute value of 
$ \Delta \omega $ is larger 
than $ \Delta \omega_0 \sim V_N ((\tau_{\rm s}/\tau_{\rm tr})(R_c/a))^{1/2}
$ or smaller than 
$ \Delta \omega'_0 \sim  V_N (\tau_{\rm s}/\tau_{\rm tr})(R_c/a)  $. 
The exact definitions of 
$ \Delta \omega_0 $ and $ \Delta \omega'_0 $ are given in (\ref{550}) and 
(\ref{555}). When $ |\Delta \omega| > \Delta \omega_0 $, we obtain 
zero contribution to the photoconductivity for both directions (\ref{550}). 
In the case $ |\Delta \omega|  <\omega'_0 $, we obtain nonzero  
contributions to the photoconductivities for the two mechanisms.   
The exact expressions for the photoconductivities 
are given in (\ref{560}), (\ref{565}), 
(\ref{570}) and (\ref{575}). All formulas are also valid for 
$ T \gg \omega_c $ which corresponds to the case of integer filling fraction, 
$ \Delta \mu =\pm V_N $. We have 
summarized in (\ref{580}) the scaling relations for the various 
photoconductivities in the striped state regime. We obtain that for both 
directions the photoconductivity of the distribution function mechanism 
is dominant over the displacement mechanism. Under the consideration of   
the conductivity expressions (\ref{570}), (\ref{575}) and Fig.~1 we obtain 
the remarkable fact that the sign of the photoconductivities of both 
directions at $ \omega \approx \omega_c $ are reversed to each other.  
We did not provide explicit conductivity expressions in this paper for the 
case $ a \lesssim R_c,  \ell_B^2/\xi $. The 
isotropic (delta-correlated) scattering regime for the striped 
state system is at the limit 
of this parameter range. 
When carrying out the calculation 
for the photoconductivities, 
we find a strong enhancement of  $ \sigma^{\rm DP}_{xx} $ and 
$\sigma^{\rm DF}_{xx} $ at $ \Delta \omega $ for large 
Landau level filling $ N $,
 fulfilling the 
conditions (\ref{590}) and (\ref{595}).

We also calculate the dark and photoconductivities for the 
case  $ \mu \gg T \gtrsim  V_N \gg \omega_c $ and 
  $a \ll R_c, \ell_B^2/\xi $. 
We obtain  expression 
(\ref{2532}) for the density of states.
In Fig.~2 we show the oscillatory part of the density of states.  
The dark conductivity 
in the direction parallel  to the modulation is given in 
(\ref{2540}), in the perpendicular direction in (\ref{2560}).  
 The various photoconductivities in the regime $ V_N \gg \omega_c $  
are given in (\ref{2590}), (\ref{2610}), (\ref{2635}) and (\ref{2660}). 
Their scaling relations  are given in 
(\ref{2665}). We obtain that the photoconductivity parallel to 
the periodic modulation is governed by the distribution 
function mechanism and oscillates as a function of $\omega$. 
This should result in 
zero resistance states in certain frequency regions 
$ \omega \approx \omega_c $. The 
photoconductivity perpendicular to the modulation direction is governed  
by the distribution function 
part when $ \omega  \gg V_N $.
In the case $ \omega  \ll V_N $, the photoconductivity 
is dominated by 
the displacement mechanism with positive non-oscillating values.
The oscillating photoconductivities all behave similarly as a 
function of $ \omega, \omega_c $. In Fig.~3, we show the  
behaviour of the various photoconductivities in 
the case $ V_N \gg \omega_c $   
    
Finally, we give in (\ref{1020}) with (\ref{1040})-(\ref{1060}) the various 
photoconductivities in the range $ \omega \gtrsim \omega_c $ 
for an arbitrary form of the microwave field. We obtain a polarization 
dependence only for   
$ \sigma_{xx}^{\rm DP} $, i.e.  in  the direction parallel to the modulation. 
This polarization dependent term is given by the last expression in 
(\ref{1040}).  \\[0.1cm]  

Beside the results mentioned above, we also obtain results 
which are interesting 
from the theoretical point of view. For calculating the conductivities 
for $ V_N \gg \omega_c $ we carried out a calculation 
of the eigenfunction and eigenvalues of an electron in a homogeneous 
magnetic field and an unidirectional periodic 
 potential in the WKB approximation. 
As is well known, the physics in high Landau levels is described correctly 
within this approximation. We obtain no corrections 
to the eigenvalues in comparison to the first order 
result in the potential amplitude $ \tilde{V} $ 
used in section III and IV and also in 
Ref. \cite{Dietel1}. The WKB eigenfunction is 
a mixing of Landau level functions. The $x $-position operator matrix
elements with respect to the WKB eigenfunctions are the same as in the first 
order in $ \tilde{V} $ approximation. 
We furthermore calculated  the expectation value of plane waves with respect 
to these WKB eigenfunctions. We obtain corrections to the first order result. 
When calculating the various 
conductivities we have to calculate the impurity averaged square 
of these matrix elements. We have shown in section V B that this 
quantity is the same as in first order perturbation theory in $ \tilde{V} $
for large Landau level index $ N $. 
Summarizing, we obtain only corrections in the  
conductivity expressions  in comparison   
to the first order formulas of section III by the fact that the  
oscillating energy bands overlap for large $ V_N / \omega_c $. 
These  corrections were taken into account in section VI.

\section{The calculation of the various conductivities for 
temperature $ {\bf T\ll V_N} $ 
and arbitrary filling fraction} 
The system under consideration is a 2D electron system in the presence of
a homogeneous magnetic field $ B $ and a periodic potential 
$ V({\bf r})={\tilde V}\cos(Qx) $ in $x$-direction 
with period $a=2\pi/ Q$. 
The system is under influence of 
an impurity potential $ U $ of correlation 
$  \langle U({\bf r})U({\bf r}')\rangle = W({\bf r}-{\bf r}')$ 
and correlation length $ \xi $. 
In Ref.\ \onlinecite{Dietel1}, 
we examined the two extreme cases of either a short range 
(delta-like) potential $ \xi \ll 1/k_F $ or a long range  
potential $ \xi \gg 1/k_F $ where all conductivity expressions 
were determined explicitly for these two limiting cases, where $\xi$ is the correlation length of the impurity potential.
By taking the full asymptotic 
expansions (\ref{a640}) for the Laguerre polynomials, one can argue that all results of Ref. \onlinecite{Dietel1} are also valid
for an arbitrary impurity potential.

In order to neglect 
quantum mechanical interference effects, we have to assume 
throughout this work that   
\begin{equation}
\xi \ll  R_c   \,.      \label{12}
\end{equation}  

At first, we consider the case of a periodic potential of an amplitude
which is small in comparison with the Landau level spacing, i.e.  $
\tilde{V} \ll \omega_c $. In this limit, we can determine physical
quantities within first order perturbation theory in $ \tilde{V} $.
The energy dispersion is given by $ \epsilon^0_{nk} \simeq \omega_c
(n+{1/2}) + V_{n}\cos(Qk\ell_B^2)$, where $ \ell_B=(\hbar/eB)^{1/2}$
denotes the magnetic length.  The eigenstates are the Landau level
wavefunctions $|nk\rangle  $ in the Landau gauge.  The amplitude $V_n$ is
given by
\begin{equation}  
  V_n = {\tilde V} \exp[-Q^2\ell_B^2/4]L_n(Q^2\ell_B^2/2) 
\approx  {\tilde V} J_0\left(2\pi \frac{R_c}{a} \right)    \,.  \label{15} 
\end{equation} 
It is useful to define the corresponding density of states (DOS) of the 
$n$th Landau level by
\begin{equation} 
\nu_n^*(\epsilon) = \nu_n^* \tilde{\nu}_n^*(\epsilon)  \,.       \label{20} 
\end{equation} 
with 
\begin{equation} 
\nu_n^*  = \nu^*= 1/2 \tilde{V} \pi^2\ell_B^2 \quad , \quad                
\tilde{\nu}_n^*(\epsilon)  =  \frac{1}{ \sqrt{1-[(\epsilon-E_n)/
\tilde{V}]^2}} 
                    \label{40}
\end{equation} 
and $ E_n=(n+1/2) \omega_c $.  
We denote the DOS of the system by $ \nu^*(\epsilon) $. It can be 
expressed  by 
\begin{equation}
\nu^*(\epsilon)=\nu^*\tilde{\nu}^*(\epsilon) = \nu^*\sum_n \tilde{\nu}_n^*(\epsilon) \,.        
   \label{45}
\end{equation}   

In the following, we first calculate the {\it dc} conductivities in
the $x$- and $y$-direction without microwave field and then the {\it dc } conductivities in the $x$- and $y$-direction when
a microwave field is applied to the sample. These conductivities were
calculated in Ref. \onlinecite{Dietel1} for arbitrary large filling
fraction and $ T \gg V_N $, $ a \ll R_c,\ell_B^2/\xi $.  In this section
we extend our analysis to the
regime $ a \ll R_c,\ell_B^2/\xi $ by considering the various
conductivities for $ T \ll V_N $.

As in Ref. \onlinecite{Dietel1}, we formulate the results 
in terms of the diffusion constants 
in the $x$- and $y$-directions 
\begin{equation} 
D_{xx}(\epsilon)  =  
\frac{R_c^2}{2\tau^*_{\rm tr}(\epsilon)} \quad \mbox{and} \quad 
D_{yy}(\epsilon)  =  v_y(\epsilon)^2 \tau^*_s(\epsilon) =
 \frac{1}{(\pi a \nu^*(\epsilon))^2} 
\tau^*_s(\epsilon) \,.  \label{103}
\end{equation}
Here $ \tau^*_s(\epsilon)=\tau_{\rm s}   \nu_0/     \nu^*(\epsilon) $ and 
$ \tau^*_{\rm tr}(\epsilon)=\tau_{\rm tr} \nu_0 /\nu^*(\epsilon) $ 
where $ \tau_{\rm s} $, $\tau_{\rm tr} $ are the single particle and 
transport scattering times, respectively, and $ \nu_0 $ is the density of states 
without unidirectional potential and without external magnetic field.
$ v_y(\epsilon) $ is the velocity of the electron with energy $ \epsilon$ 
in the $y$-direction. We define $ \tau^*$ by $ \tau^*(E_n)$ where  $ \tau^* $ 
stands for the various transport times. 

In the following, we calculate the dark conductivities in subsection 
A and the microwave induced photoconductivities in subsection B.

\subsection{The dark conductivities for $ {\bf T \ll V_N} $, $ {\bf  
a \ll R_c, 
\ell_B^2/\xi }$} 

The dark conductivities in $x$- and $y$-direction are given by \cite{Dietel1} 
\begin{eqnarray}
    \sigma_{xx,yy} = 
\int d\epsilon 
\left(-\frac{\partial f(\epsilon)}{ \partial \epsilon}\right)
      \sigma_{xx,yy}(\epsilon)                   \label{105}
\end{eqnarray}
with the one particle distribution function 
$ f(\epsilon)=1/(1+\exp[-\beta(\epsilon-\mu)]) $. 
The conductivity per energy is given by   
\begin{eqnarray}
    \sigma_{xx,yy}(\epsilon) = e^2 D_{xx,yy}(\epsilon)\nu^*(\epsilon).
\label{110}
\end{eqnarray}
 These expressions were derived explicitly without temperature 
restrictions in Ref. \onlinecite{Dietel1}.  
\subsection{The photoconductivities for $ {\bf T \ll V_N} $, $ {\bf  
a \ll R_c}, 
{\bf \ell_B^2/\xi }$ }  

We now come to the discussion of the photoconductivities 
of both mechanisms 
\subsubsection{Photoconductivities of the displacement mechanism}

For the conductivity in the $x$-direction $  \sigma_{xx}^{\rm DP} $ 
we find a strong dependence of the $ T \ll V_N $ result 
on the special filling fraction. 
By carrying out a similar calculation as in 
Ref.\ \onlinecite{Dietel1} we obtain 
\begin{eqnarray}
  \sigma_{xx}^{\rm DP} \backsimeq  \pi^2 \left[e^2D_{xx}\nu^* \right] 
\frac{\tau_{\rm s}^*}{\tau^*_{\rm tr}}  
    \left({eER_c\over 4\Delta\omega}\right)^2 
A_1\left(\frac{\Delta \omega}{ 2V_N},
\frac{\Delta \mu}{V_N}\right)  
\label{140}
\end{eqnarray}
with the function
\begin{equation}
    A_1\left(\frac{\Delta \omega}{ 2V_N},
\frac{\Delta \mu}{V_N}\right)=   
-\frac{6}{  \pi^2}\frac{\partial}{\partial \Delta \omega }    
   \sum_{\sigma \in \{\pm\}} 
 \int\limits_{E_N+\sigma \Delta \mu}^{E_N+V_N }   
\, d\epsilon \; \tilde{\nu}^* (\epsilon) 
\tilde{\nu}^*(\epsilon+\Delta \omega)\,.      \label{150} 
\end{equation}
and $ \Delta  \mu =\mu-(N+1/2) \omega_c $, 
$ \Delta \omega =\omega_c-\omega$. The exact numerical prefactor 
in  (\ref{140}) depends on the details of disorder. For  
delta correlated impurities expression (\ref{140}) is exact \cite{Rem1}.  
The integral in (\ref{150}) is of incomplete elliptical type.  
It is easily seen that one gets  $ A_1(x,\pm 1)=
-(3/ \pi^2) (\partial/\partial x) K(\sqrt{1-x^2})$, 
which corresponds to a system of integer filling. 
We obtain for integer filling the same result for the conductivity  as for 
 $ T \gg V_N $ \cite{Dietel1}.  \\
 
Next, we calculate the displacement current in the $y$-direction.
This was done in Ref. \onlinecite{Dietel1} by calculating the eigenstates 
for electrons for a modulation potential in $x$-direction 
and a small additional {\it dc} field in $y$-direction, neglecting 
Landau level mixing. These  eigenstates $ |n {\cal E}_{l}\rangle $ 
are the dual to the basis states $ |n k\rangle  $ in a similar 
manner as the basis of plane waves are the Fourier dual to the position 
eigenstates. By representing  the ground state wavefunction as a linear 
combination of a Slater determinants of the functions 
$ |n {\cal E}_{l}\rangle  $, 
we do not get a single term, which means that one 
cannot apply  Fermi's golden rule formula 
of \cite{Dietel1} to calculate the conductivities. 
By a generalization,  we show in appendix A 
that one can nevertheless apply the simple Fermi's golden rule formula 
given in (\ref{a40}) 
when using for the distribution function in this formula 
\begin{equation}
f^y_{n l}=\frac{2 \pi \ell_B^2}{(L_x L_y)} 
\sum_k f(\epsilon_{nk})       \label{155}
\end{equation} 
where $ f^y_{nl} $
is the value of the distribution function for the state 
$ |n {\cal E}_{l} \rangle $. 
Thus, we obtain that $ f^y_{nl} $ depends only on the Landau 
level index $n$. The reason for this is 
that the absolute value of the overlap of $ |n {\cal E}_{l}\rangle $ and
$ |n k\rangle  $ does not depend on $k $ and $l $. 
From this, we find that the conductivity $  \sigma_{yy}^{\rm DP} $ 
 is independent of the filling fraction 
\begin{equation}
  \sigma_{yy}^{\rm DP} =
 \left[ e^2D_{yy}\nu^*\right]\left(aB/\pi\sqrt{\tau_{\rm s}^*\tau^*_{\rm tr}}
  \over E_{\rm dc} \right)^2
  \left(\frac{eER_c}{   4\Delta \omega}\right)^2
  A_2\left(\frac{\Delta \omega}{ 2V_N},
\frac{\Delta \mu}{V_N}\right)                        \label{160}
\end{equation}
with 
\begin{equation}
   A_2\left(\frac{\Delta \omega}{ 2V_N},
\frac{\Delta \mu}{V_N}\right) = \frac{\Delta \omega}{V_N^2} 
\int\limits_{E_N-V_N}^{E_N+V_N}  \, d\epsilon \; \tilde{\nu}^*(\epsilon) 
\tilde{\nu}^*(\epsilon+\Delta \omega). \label{170}
\end{equation}
This expression was also obtained for $ T \gg V_N $ in Ref. 
\onlinecite{Dietel1}. 
By carrying out the integral in $ A_2 $, we obtain $ 
 A_2(x,y)
=2x  K(\sqrt{1- x^2}) $.
Note the singular dependence of the displacement contribution 
to the transverse photoconductivity on the {\it dc} electric field 
$E_{\rm dc}$. 
This singularity is 
cut off for small {\it dc} 
 electric fields by inelastic processes \cite{Dietel1}
when $E_{\rm dc} \sim 
E^*_{\rm dc}$, where $E_{\rm dc}^*= 
Ba/2\pi\sqrt{\tau_{\rm in}\tau_{\rm s}^*}$.
For $E_{\rm dc}\ll E_{\rm dc}^*$, the photoconductivity crosses over to
Ohmic behavior, matching with 
Eq. (\ref{160}) for $E_{\rm dc} \sim E^*_{\rm dc}$.\\

\subsubsection{
Photoconductivities  of the  distribution function mechanism}  
First, we calculate the photoconductivity $ \sigma_{xx}^{\rm DF} $ in
the $ x$-direction.  The calculation of Ref. \onlinecite{Dietel1} can
be easily transferred to the case $ T \ll V_N $.  Due to the
irradiation of the system by the microwave field, the distribution
function is changed from its dark value. We obtain
 \begin{equation}
\label{176}
        \delta f(\epsilon) \approx  2 \frac{\tau_{\rm in}}{\tau_{\rm tr}}
 \left(\frac{eER_c}{ 4\Delta\omega}\right)^2
\sum_{\sigma=\pm}[ f(\epsilon-\sigma \omega) - f(\epsilon)] \;  
\frac{\nu^{*}(\epsilon+\sigma\Delta\omega)}{\nu_0}          \;.
\end{equation} 
By using this distribution function we obtain 
 \begin{eqnarray}
\label{178}
\sigma^{\rm DF}_{xx}= 4 
 \left(\frac{\tau_{\rm in} }{ 
\tau_{\rm tr}^*} \right)\left({eER_c\over 4\Delta\omega}\right)^2 
\left[e^2D_{xx}\nu^*\right] B_1\left(\frac{\Delta \omega}{ 2V_N},
\frac{\Delta \mu}{V_N}\right)
\end{eqnarray}
with 
\begin{equation}
   B_1\left(\frac{\Delta \omega}{ 2V_N},
\frac{\Delta \mu}{V_N}\right) =      
 - \frac{1}{2}
\sum_{\sigma \in \{\pm\}}
\left[\left( \frac{\partial}{\partial \Delta \omega}-
\sigma \frac{\partial}{\partial \Delta \mu} \right) 
\left(\int\limits_{E_N+\sigma \Delta \mu}^{E_N+V_N}  d\epsilon 
 + 
\int\limits_{E_N-V_N}^{E_N-\Delta \omega' +\sigma \Delta \mu} d\epsilon\right)    
\, \; (\tilde{\nu}^*(\epsilon))^2 
\tilde{\nu}^*(\epsilon+\Delta \omega) \right]_{\Delta \omega'=\Delta \omega}
\;. \label{180}   
\end{equation} 
This results in 
\begin{equation}
B_1(x,y) \approx
 Z[x,y] {1\over 16}{1-2|x|\over (|x|-|x|^2)^{3/2}}
\ln \left({V_N\over \Delta }\right) \,  {\rm sgn \,x}  \,.               \label{182}
\end{equation}
$ \Delta $ is an effective 
energy cutoff at the band edges resulting
from an additional energy broadening  
for example due to impurity scattering or a finite {\it dc} field. The 
filling fraction dependent prefactor $ Z[x,y] $ 
is given by $ Z[\Delta\omega/ 2V,\Delta \mu/V_N]=1 $ 
for $ T \gg V_N $.
For $ T \ll V_N $ and $ \Delta \omega > 0 $ 
we have
\begin{equation} 
 Z\left[\frac{\Delta \omega}{ 2V_N}, \frac{\Delta \mu}{V_N}\right] = 
\left\{ 
 \begin{array}{ccc}
 1 &\qquad  \mbox{for} \qquad & \Delta \omega \le 
V_N-|\Delta \mu|    \\  
 \frac{1}{2} & \mbox{for} & V_N -|\Delta \mu| \le \Delta \omega 
\le V_N +|\Delta \mu|
\\
0  &  \mbox{for} & \Delta \omega \ge V_N+|\Delta \mu|  \,.           
\end{array} \right.                   \label{185}
\end{equation}   
For $ T \ll V_N $ and $\Delta  \omega < 0 $ we have 
$ Z[\Delta\omega/ 2V_N,\Delta \mu/V_N]=2-Z[-\Delta\omega/ 2V_N,\Delta \mu/V_N] $. The zero in the last line in (\ref{185}) 
reflects the fact that the conductivity is not determined by the logarithm 
of the broadening energy cutoff parameter $ \Delta $ 
which is assumed to be small compared to $V_N$.   \\

Finally, we calculate the {\it dc} current in the $y$-direction coming
from the distribution function mechanism.  One can immediately
transfer the calculation of Ref.\ \onlinecite{Dietel1} for $
\sigma_{yy}^{\rm DF} $ to the case $ T \ll V_N $ which yields
\begin{equation}
\label{190}
    \sigma_{yy}^{\rm DF}=  4 
\left(\tau_{\rm in}\over\tau_{\rm tr}^*\right)
    \left({eER_c\over  4\Delta\omega}\right)^2[e^2 D_{yy} \nu^*] 
B_2\left(\frac{\Delta \omega}{ 2V_N},
\frac{\Delta \mu}{V_N}\right) 
\end{equation}
with 
\begin{equation}
B_2\left(\frac{\Delta \omega}{ 2V_N},
\frac{\Delta \mu}{V_N}\right)= 
\frac{1}{2} 
\sum_{\sigma \in \{\pm\}} 
\left(\int\limits_{E_N+\sigma \Delta \mu}^{E_N+V_N} d \epsilon  
 + \int\limits_{E_N-V_N}^{E_N-\Delta \omega +\sigma \Delta \mu} d\epsilon \right)    
\left(\frac{\partial }{\partial \epsilon}
\frac{1}{(\tilde{\nu}^*(\epsilon))^2} \right)
 \tilde{\nu}^*(\epsilon+\Delta \omega)     \,.     \label{195}
\end{equation}
This integral can be expressed as a sum of elementary functions 
\cite{Dietel1} which will not be stated due to its length. 
In the special case of integer filling (or $ T \gg \omega_c $) 
\cite{Dietel1} we obtain
$ B_2(\Delta \omega/2V_N,\pm 1) =[4 |x|(\arcsin(1-2|x|)+\pi/2)-
4\sqrt{|x|-|x|^2}] {\rm sgn}\,  x $ \cite{Rem1}.  

By taking into account these calculations we obtain the following 
scaling relations for the various photoconductivities
\begin{equation}
  \sigma_{xx}^{\rm DF} \sim 
\left(\frac{\tau_{\rm in}}{\tau^*_s} \right)  \sigma_{xx}^{\rm DP} \quad , 
\quad    \sigma_{yy}^{\rm DF} 
\sim \sigma_{yy}^{\rm DP}\,.   \label{200}
\end{equation}
Summarizing, we obtain a photoconductivity 
 in the direction parallel to the modulation which is governed 
by the distribution function mechanism. 
In the perpendicular direction  
both mechanisms  contribute equally. This was derived before in 
\cite{Dietel1} for $ T \gg V_N $. In the case of the 
photoconductivity experiments 
without the periodic potential the distribution function mechanism 
is the dominant contribution to the photoconductivity \cite{Dmitriev2}. 
We point out that 
the photoconductivities  of integer 
filling fraction $ \Delta \mu = \pm V_N $  for $ T \ll V_N $ 
are in accordance with the photoconductivities 
for   $ T \gg V_N $ calculated in \cite{Dietel1} which are odd  
functions of $ \Delta \omega $. 
This is except for $ \sigma_{yy}^{\rm DP} $ not the case 
for systems of fractional filling and $ T \ll V_N $. We show in Fig.~1 
the functions $ A_i(\Delta \omega/2V_N ,0) $ and 
$B_i(\Delta \omega/2V_N,0) $ corresponding to upper Landau level 
half filling fraction 
most relevant for the striped state.
Note the singularity of $ A_1(\Delta \omega/2V_N ,0) $ at 
$ \Delta \omega=\pm V_N  $ coming from the singularity of the density 
of states (\ref{40}) at the band edge.   

\begin{figure}
 \begin{center}
 \psfrag{x} 
 {\scriptsize{\hspace*{-0.5cm} $\frac{\Delta \omega}{2 V_N} $}}
 \psfrag{A1} 
 {\scriptsize{$A_1 $}}
\psfrag{B1} 
 {\scriptsize{$B_1/{\rm ln}(V/\Delta)$}}
 \psfrag{A2} 
 {\scriptsize{$A_2 $}}
\psfrag{B2} 
 {\scriptsize{$B_2 $}}
\includegraphics[height=6cm,width=12cm]{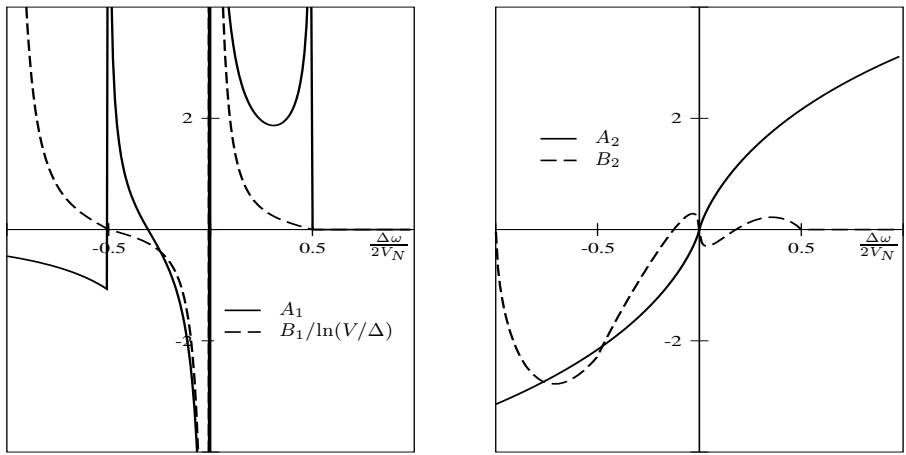}
 \end{center}
 \caption{The functions $ A_i(\Delta \omega/2V_N ,0) $ (solid  line) given by 
(\ref{150}), (\ref{170})  and 
$ B_i(\Delta \omega/2V_N ,0) $ (dashed line) given by (\ref{182}), 
(\ref{195}) 
describing the dependence of the 
conductivities on the microwave frequency for the upper Landau level 
half filled system 
$ \Delta \mu =0 $.}
\end{figure}

Summarizing, we obtain singularities of the photoconductivities 
at $ \Delta \omega =0 $ but also at some other frequencies for 
$ \sigma^{\rm DP}_{xx} $ and  $ \sigma^{\rm DF}_{xx} $. 
In realistic systems these photoconductivities are cut by impurity
scattering. The origin of the singularities comes from the prefactors 
$ (e E R_c/ 4 \Delta \omega)^2 $ in the photoconductivity 
expressions (\ref{140}), (\ref{160}), (\ref{178}), (\ref{190}) resulting 
in the divergence of the photoconductivities at $ \Delta \omega =0 $    
and also from the singularities in the 
density of states at the band edge resulting in the  
singularities in   
$ A_1$ and $ B_1 $. As was shown explicitly 
in Ref . \onlinecite{Dmitriev1} 
for the distribution function mechanism of  the 
system without modulation potential,   
the $  (e E R_c/ 4 \Delta \omega)^2 $ term should be 
cut at $ |\Delta \omega |\sim 1/ \tau^*_{\rm tr} $. The cut off value  
due to corrections of the density of states at the 
band edge is equal to the  
Landau level broadening value of  the system without modulation given by  
$ |\Delta \omega-\Delta \omega_s| 
\sim  \sqrt{\omega_c/\tau_{\rm s }} $ for a singularity 
at $\Delta \omega = \Delta \omega_s  $. We point out 
that a proper derivation of the form of the conductivity expressions 
at the various singularities requires a calculation of 
 higher order corrections to the transition amplitude
including vertex corrections which is out of the scope of this work.

\section{Calculation of  conductivities 
 without restrictions on  modulation wavelength and  cyclotron radius}

There are different types of ground states for the various filling
fractions of a 2DEG in a homogeneous magnetic field. One
of the most prominent is relevant for filling fraction $2 \lesssim N
\lesssim 8 $.  This state consists of a charge density wave, which can
be shown to be the ground state within Hartree-Fock approximation
\cite{Moessner, Fogler}.  The wavelength of the density wave is
approximately given by $ a \approx 2.6 R_c $, i.e. out of the limits
considered up to now. In this section, we therefore extend our
previous considerations to the case without restrictions to $ a $ and
$ R_c $.  We believe that our model with the additional unidirectional
periodic modulation can be used for the striped state system because
when restricting the Hartree as well as the Fock potentials to the
upper Landau level, both potentials have the same form \cite{Fogler}.
Since the unidirectional periodic potential is of Hartree form, we
believe that we can use this potential also for modelling the Fock
potential by adjusting the potential amplitude $ \tilde{V} $ to the
full amplitude of the Hartree-Fock potential of the striped state.

In the following two subsections, we discuss the dark and the photoconductivities for the  case
$ {\rm min}[1/\xi,2 k_F]  \ll  a/\ell_B^2 $ which is equivalent to 
$ \sqrt{\tau_{\rm tr}/\tau_{\rm s}} \, (a/R_c) \gg 1 $.
This case is relevant for the striped state if scattering takes place only between nearest neighbor stripes.
In the last subsection, we discuss the 
various conductivities in the transition region for smooth disorder, i.e.  
$ {\rm min}[1/\xi,2 k_F]  \gtrsim  a/\ell_B^2 $ or 
$ \sqrt{\tau_{\rm tr}/\tau_{\rm s}} \, (a/R_c) \lesssim 1 $, respectively.  
Throughout this section, we redefine $ 1/\xi $ to be 
the minimum of the inverse of the correlation length and  $  2 k_F $.

\subsection{Calculation of the dark conductivities in the case 
$ {\bf  \sqrt{\tau_{\rm tr}/\tau_{\rm s}} \,(a/R_c) \gg 1 } $}
 
In general, the dark conductivities $ \sigma_{xx} $, 
$ \sigma_{yy} $  can be calculated using Eq. (\ref{105}) and Eq.  (\ref{110}) by replacing the scattering times  $ \tau_{\rm s} $ 
and $ \tau_{\rm tr} $ in the diffusion constants, Eq. (\ref{103}),
by the energy dependent 
scattering times  $ \tilde{\tau}_s(\epsilon) $ and 
$ \tilde{\tau}_{\rm tr1}(\epsilon)  $
which are defined by 
\begin{align} 
& \frac{1}{\tilde{\tau}_s(\epsilon)} 
=  a \nu_0 \sum_{j \in \mathbb{Z}} 
   \int \,dq_x  e^{-\frac{q'^2_j\ell_B^2}{ 2}} 
[L_n({q'^2_j\ell_B^2\over 2})]^2
\tilde{W}(q'_j)      
    \nonumber \\
& 
\frac{1}{\tilde{\tau}_{\rm tr1} (\epsilon)} = 
 \frac{a \nu_0}{4N} \sum_{j \in \mathbb{Z}} 
   \int \,dq_x \, e^{-\frac{q^2_j\ell_B^2}{ 2}} 
\left(\frac{j a}{\ell_B}\right)^{2}        
 [L_n({q_{j}^2\ell_B^2\over 2})]^2
\tilde{W}(q_j)+e^{-\frac{q'^2_j\ell_B^2}{ 2}} 
\left(\frac{2 k \ell_B^2 + j a }{ \ell_B}\right)^2
[L_n({q'^2_j\ell_B^2\over 2})]^2
\tilde{W}(q'_j)    \,.   \label{510}
\end{align}
with $ {\bf q}_j=q_x {\bf e}_x + (j a/\ell_B^2){\bf e}_y  $,  
$ {\bf q}'_j =q_x {\bf e}_x  +  (2 k+ja/\ell_B^2) 
{\bf e}_y  $ and 
\begin{equation} 
k=\arccos[(\epsilon-(N+1/2) \omega_c)/V_n]/ Q \ell_B^2 .   \label{515}
\end{equation}
Here $ k \ell_B^2 $  
and $ {\bf q}_j  \ell^2_B $, $ {\bf q}'_j  \ell^2_B $ 
in (\ref{510}), (\ref{515}) can be interpreted in the following way: 
The value $ k \ell_B^2 $ corresponds to the real space position of the 
electron for which we calculate the transition rate. 
$ {\bf e}_z \times {\bf q}_j  \ell^2_B $, $ {\bf e}_z \times 
{\bf q}_j'  \ell^2_B $  
correspond to the displacement vectors of the transition. 
It is immediately seen that $ 1/\tilde{\tau}_s(\epsilon)= 
1/\tau_{\rm s} $ and $ 1/\tilde{\tau}_{\rm tr1}(\epsilon)= 1/\tau_{\rm tr} $
for $ a \ll \ell_B^2/\xi $ \cite{Dietel1}. 

For $  \xi \gg \ell_B^2/a$, we can neglect the 
first term in (\ref{510}) and  restrict the 
$ j $-sum in (\ref{510}) and (\ref{515}) only to the summand 
with $| (2 k+j a/\ell_B^2 )| \le a/2 \ell_B^2 $. 
The conductivities can then no longer be expressed in terms of the single particle and transport time of a given impurity correlation function $ \tilde{W} $.
We therefore carry out a scaling analysis of the conductivities.
Nevertheless, we obtain from (\ref{510}) the exact relation for 
$ \tilde{\tau}_s(\epsilon) $ and $ \tilde{\tau}_{\rm tr1}(\epsilon) $ 
\begin{equation}
\frac{1}{\tilde{\tau}_{\rm tr1}(\epsilon)} = 
\frac{\Delta^2_{\rm ph}}{2 R_c^2}  \frac{1}{\tilde{\tau}_{\rm s}(\epsilon)}
\,.  
 \label{518}
\end{equation}
Here $ \Delta_{\rm ph} =\rm{min}_j[|2k\ell_B^2+ja|] $ is the smallest possible displacement of an electron jump, which, in the striped state,
can be interpreted as 
the smallest distance between an electron and a hole stripe.  
By using the asymptotics of the Laguerre 
polynomials, given in App. C, we obtain (with  
$ \xi \sim \sqrt{\tau_{\rm tr}/\tau_{\rm s}} \, \ell_B^2 /R_c $)     
\begin{equation} 
\frac{1}{\tilde{\tau}_{\rm s} (\epsilon)}  \sim \left\{ 
\begin{array}{ccc}  
\frac{\xi a}{\ell_B^2 \tau_{\rm s}}
 \sim  \sqrt{\frac{\tau_{\rm tr}}{\tau_{\rm s}}} \, 
\left(\frac{a}{R_c} \right) \frac{1}{ \tau_{\rm s}} & \,  
\mbox{for $n$ fulfilling } \, & V_N - |\epsilon-(n+1/2)\omega_c|
< V_N\left(1-\cos\left( 2 \pi \frac{\ell_B^2}{2 a \xi}\right) \right) 
\sim 
V_N \frac{\tau_s}{\tau_{\rm tr}} \frac{R_c^2}{a^2}      \\ 
0 & \quad \mbox{otherwise}\,.
\end{array}  
\right. 
\label{520}
\end{equation}
Summarizing, we have 
\begin{eqnarray} 
\sigma_{xx}(\epsilon) & =&  e^2 D_{xx}(\epsilon) \nu^*(\epsilon)  
\label{523}
\frac{\tau_{\rm tr}}{\tilde{\tau}_{\rm tr1}(\epsilon)}
 \,,   \\
\sigma_{yy}(\epsilon) & = & e^2 D_{yy}(\epsilon) \nu^*(\epsilon)
\left[ \frac{\tilde{\tau}_{\rm s}(\epsilon)}{\tau_{\rm s}} \, 
\, \Theta\left(\frac{1}{\tilde{\tau}_{\rm s} (\epsilon)}\right)+
\frac{\tau_{\rm in}}{\tau^*_{\rm s}(\epsilon)} \, 
\, \left(1- \Theta
\left(\frac{1}{\tilde{\tau}_{\rm s} (\epsilon)}\right)\right)\right]
  \,.  
\label{526}
\end{eqnarray}
The values of $ \Theta(1/\tilde{\tau}_{\rm s,tr1} (\epsilon)) $ 
are  defined 
by unity in the case 
that $ 1/\tilde{\tau}_{\rm s,tr1} (\epsilon)>0 $ and zero 
otherwise.   
The dark conductivities are then calculated by the insertion 
of (\ref{523}) and (\ref{526}) in (\ref{105}). 
Thus, the dark conductivity  $ \sigma_{xx} $ can be  
calculated by the substitution of $ \tilde{\tau}_{\rm tr1}(\epsilon) $  
for $ \tau_{\rm tr} $  in (\ref{110}) \cite{Dietel1}. 
This is not the case for $ \sigma_{yy} $. 
In this case $ \sigma_{yy}(\epsilon) $ would get singular values for 
some energies $ \epsilon $ 
when we formally substitute $ \tilde{\tau}_{\rm s}(\epsilon) $ 
for $ \tau_{s} $ in (\ref{110}). 

From Eq. (\ref{526}), it is obvious that, due to the Theta function, there are two fundamentally different energy regions for the conductivity.
For the energy region where $ 1/ \tilde{\tau}_{\rm s}(\epsilon) =0 $, 
the conductivity is mainly due to inelastic scattering. In contrast, for 
the region where $1/ \tilde{\tau}_{\rm s}(\epsilon) \not =0 $,
the dominant process is the scattering by impurities. 
In order to derive the conductivity formula (\ref{526}), one 
has to solve the Boltzmann equation containing also the inelastic 
scattering time as sketched in 
App. B of Ref. \onlinecite{Dietel1}. 
For equation (\ref{526}) to be valid, the condition 
$ E_{\it dc} \ll  aB/ \tau_{\rm in} $ has to be satisfied. 
This can be shown by taking 
into account that the change in energy within the relaxation time 
$ \tau_{\rm in } $ should be smaller than the effective 
modulation potential \cite{Dietel1}. In the regime $ 
 aB/ \tau_{\rm in} \ll  E_{\rm dc} \ll 
 aB/ \sqrt{\tau_{\rm in} \tilde{\tau}^*_{\rm s}} $, where 
$ \tilde{\tau}_{\rm s} $ 
is the minimum of $ \tilde{\tau}_{\rm _s}(\epsilon) $ on the Landau band,  
we obtain an expression similar to Eq. (\ref{526}) where the second term 
on the right hand side of (\ref{526}) in the conductivity formula 
(\ref{105}) gives the result only up to numbers with respect to the exact expression. For $   aB/ \sqrt{\tau_{\rm in} \tilde{\tau}^*_{\rm s}} 
\ll  E_{\rm dc}$  
inelastic scattering  due to heating is dominant, 
which results in the behavior that 
$ \sigma_{yy} \sim (aB/E_{\rm dc}
 \sqrt{\tau_{\rm in} \tilde{\tau}^*_{\rm s}})^2 \, 
e^2 D_{yy} \nu^* $. This was shown explicitly in \cite{Dietel1}.    

By use of Eqs. (\ref{518}), (\ref{523}), (\ref{526}),  
we obtain the relation 
\begin{equation}
\sigma_{xx}(\epsilon) \, \sigma_{yy}(\epsilon) = e^4 \, 
\frac{ \Delta^2_{\rm ph}}{4 a^2 \pi^2} \,  
\Theta\left(\frac{1}{\tilde{\tau}_{\rm s,tr1} (\epsilon)}\right) \,.  
\label{528}
\end{equation}   
We point out that Eq. (\ref{528}) is an exact relation within our model
although Eqs. (\ref{523}) and (\ref{526}) with Eqs. (\ref{518}) and (\ref{520}) 
are only scaling relations.
This is due to the exact relation (\ref{518}) which 
implies a universal value for $ \tilde{\tau}_{\rm s}/ 
\tilde{\tau}_{\rm tr1}$ independent of the specific impurity configuration.   
In the case of half filling, i.e.  
$ \Delta_{\rm ph}=a/2  $, and assuming  
that the correlation length allows jumps between nearby electron and
hole stripes, i.e.   
$ \Theta(1/\tilde{\tau}_{\rm s,tr1}(\epsilon))=1 $,  we obtain in 
Eq. (\ref{528}) the 
well known semicircle law for the 
conductivities of the striped state of half integer 
filling. This rule was derived before  
in Ref.\ \onlinecite{MacDonald2} without taking into account the microscopic 
scattering process.

\subsection{Calculation of  photoconductivities for 
$ {\bf  \sqrt{\tau_{\rm tr}/\tau_{\rm s}} \, (a/R_c) \gg 1}  $}

As in the calculation of the 
dark conductivities above, we have to replace the 
scattering times in the integrals of section III by energy dependent 
scattering times in order to obtain the photoconductivities $ \sigma^{\rm photo} $ in the range 
$ \xi \gg \ell_B^2/a  $. 
\begin{eqnarray} 
\frac{1}{\tilde{\tau}_{\rm tr2} (\epsilon)} \nts{2} & = & \nts{2} 
 \frac{a \nu_0}{4} \sum_{j \in \mathbb{Z}} 
   \int \,dq_y 
\;e^{-\frac{q^2_j\ell_B^2}{ 2}} \left[L_N({q_{j}^2\ell_B^2\over 2})-
L_{N-1}({q_{j}^2\ell_B^2\over 2})\right]^2\;
\tilde{W}(q_j)+e^{-\frac{q'^2_j\ell_B^2}{ 2}}
\left[L_N({q_{j}'^2\ell_B^2\over 2})-
L_{N-1}({q_{j}'^2\ell_B^2\over 2})\right]^2\;
\tilde{W}(q'_j)  \,,   \nonumber   \\
\frac{1}{\tilde{\tau}_{\rm 2} (\epsilon)} & = &  
 \frac{a \nu_0}{12N} \sum_{j \in \mathbb{Z}} 
   \int \,dq_y \;  e^{-\frac{q^2_j\ell_B^2}{ 2}}      
\left(\frac{j a+\Delta a}{\ell_B}\right)^{2}  
\left[L_N({q_{j}^2\ell_B^2\over 2})-
L_{N-1}({q_{j}^2\ell_B^2\over 2})\right]^2\;
\tilde{W}(q_j)    \nonumber \\
& & 
+e^{-\frac{q'^2_j\ell_B^2}{ 2}} \left(\frac{2 k \ell_B^2 + j a -\Delta a }{ \ell_B}\right)^{2} 
\left[L_N({q_{j}'^2\ell_B^2\over 2})-
L_{N-1}({q_{j}'^2\ell_B^2\over 2})\right]^2 \;
\tilde{W}(q'_j)  \,.  \label{530}
\end{eqnarray}
with $ {\bf q}_j=q_x {\bf e}_x + (j a+\Delta a)/\ell_B^2){\bf e}_y  $,  
$ {\bf q}'_j =q_x {\bf e}_x 
 + (2 k+(j a - \Delta a)/\ell_B^2 ){\bf e}_y  $ and $ k\ell_B^2 $ from (\ref{515}).  

The value $ \Delta a $ is defined by the condition for energy  
conservation  
\begin{equation} 
\cos\left(2 \pi \frac{k \ell_B^2 +\Delta a}{a}\right)-
\cos\left(2 \pi \frac{k \ell_B^2}{a}\right)+\frac{\Delta \omega}{V_N}=0 \,.
\label{535}
\end{equation}
This equation reflects the fact that electrons can only be scattered 
to positions of energy difference 
$ \omega $ under the combined influence  
of impurity scattering and microwave irradiation.  
The replacement of the scattering times by their energy-dependent values amounts to the following substitutions 
\begin{equation}
\begin{array}{lcclcc}
\tau_{\rm s} /\tau^2_{\rm tr} \rightarrow 
1/\tilde{\tau}_2(\epsilon) &  {\rm  in } &   
(\ref{140}), (\ref{150})\;  ; & 1/\tau_{\rm tr}
\rightarrow 1/\tilde{\tau}_{\rm tr2}(\epsilon) &  {\rm in} &   
 (\ref{160}), (\ref{170});
\\ 
 1/\tau^2_{\rm tr}
\rightarrow 1/\tilde{\tau}_{\rm tr1}(\epsilon) 
\tilde{\tau}_{\rm tr2}(\epsilon) &  {\rm  in} &    
(\ref{178}), (\ref{180}) \; ; & 
\tau_{\rm s} /\tau_{\rm tr} \rightarrow \tilde{\tau}_s(\epsilon)/
\tilde{\tau}_{\rm tr2}(\epsilon) &  {\rm  in } &    (\ref{190}), (\ref{195}).
\end{array}
\label{538}
\end{equation}   
We point out that the derivative in (\ref{195}) has to 
be applied to the energy 
dependent scattering term 
$ \tilde{\tau}_{s}(\epsilon) $ in $ D_{yy}$ (\ref{190})
but not to $ \tilde{\tau}_{\rm tr2}(\epsilon) $. 
Similarly as in the calculation of the dark conductivity (\ref{518}), we 
have an exact relation between $ 1/\tilde{\tau}_{\rm tr2} $ and 
$ 1/\tilde{\tau}_{\rm 2} $. This can be calculated from  
(\ref{530}) resulting in 
\begin{equation}
\frac{1}{\tilde{\tau}_{\rm 2}(\epsilon)} = 
\frac{2}{3}
\frac{(\Delta a)^2}{R_c^2}   
\frac{1}{\tilde{\tau}_{\rm tr2}(\epsilon)} \sim \frac{a^2}{R_c^2} 
\frac{(\Delta \omega \,\tilde{\nu}^*(\epsilon))^2}{V^2_N} \frac{1}{\tilde{\tau}_{\rm tr2}(\epsilon)}.
  \label{539}
\end{equation}
The scaling law at the right hand side 
of Eq. (\ref{539}) can be derived by similar considerations 
as in Eq. (\ref{550}) below.
By using the asymptotics of the Laguerre polynomials from App.~C we obtain 
the following scaling law for $ \tilde{\tau}_{\rm tr2} $
\begin{equation} 
\frac{1}{\tilde{\tau}_{\rm tr2} (\epsilon)} 
\qquad \sim \qquad \left\{ 
\begin{array}{ccc}  
\frac{a \xi }{\ell_B^2 \tau_{\rm tr}} \sim   
\sqrt{\frac{\tau_{\rm tr}}{\tau_{\rm s}}} \, 
\left(\frac{a}{R_c} \right) \frac{1}{\tau_{\rm tr}}
 & \, \mbox{for some n with} 
\,  & 
\left|\arccos\left[\frac{\epsilon-\Delta \omega}{V_n}\right]- 
\arccos\left(\frac{\epsilon}{V_n}\right)\right| < 
\frac{(2 \pi) \ell_B^2}{a \xi} 
  \,, 
\vspace*{0.1cm} \\ 
0 & \qquad \mbox{otherwise} \,. 
\qquad & 
\end{array}  
\right.
\label{540}
\end{equation}
From Eq. (\ref{540}), we obtain $ 
1/\tilde{\tau}_{\rm tr2;2} (\epsilon) =0 $ for  
$ \Delta \omega >  2 V_{N}\sin[(2\pi) \ell_B^2/ 2 a \xi] $.   
This results in the following expression for the photoconductivities 
\begin{eqnarray}
\sigma^{\rm photo}=0       \qquad \mbox{for} 
\qquad  |\Delta \omega| >  \Delta \omega_0 
& := &  2V_N 
\sin\left[\frac{(2\pi) \ell_B^2}{2 a \xi}\right] 
\approx   V_N\left( \frac{2 \pi \ell_B^2}{ a \xi}\right) 
      \sim V_N \sqrt{\frac{\tau_{\rm s}}{\tau_{\rm tr}}} \, 
\left(\frac{R_c}{a}\right)              \,.
 \label{550}
\end{eqnarray}
Here $ \sigma^{\rm photo} $ stands for $ \sigma^{\rm DP}_{xx} $, 
$ \sigma^{\rm DP}_{yy} $, $ \sigma^{\rm DF}_{xx} $ or 
$ \sigma^{\rm DF}_{yy} $. 
The fact
that we obtain zero photoconductivity for 
$ |\Delta \omega| > \Delta  \omega_0 $ can be understood by taking into 
account that the microwave-mediated  impurity jump of an electron 
has to fulfill an energy conservation condition which restricts the 
number of possible final states. 
The momentum transfer of the impurity to the 
electron is further restricted by the maximal value $ 1/\xi $.  
When taking into account that the momentum $ k $ of the electron and its 
$X $ coordinate are related via $ X =k\ell B^2 $ we obtain 
Eq. (\ref{550}). 
By using Eq. (\ref{540}) 
we are now able to calculate the various scaling laws 
for the photoconductivities of section III for the case
$ |\Delta \omega| \ll  \Delta \omega_0 $.  
 
\subsubsection{Photoconductivities for the displacement mechanism}

By using Eqs. (\ref{140}), (\ref{160}) and (\ref{539}) we obtain for the 
conductivity in the regime 
\begin{equation}
|\Delta \omega| < \Delta\omega'_0= 
V_{N}\left(1- \cos\left(\frac{\pi \ell_B^2}{ a \xi}\right)\right)
\approx   \frac{1}{2} V_N\left( \frac{\pi \ell_B^2}{ a \xi}\right)^2 \sim 
 V_N \left(\frac{\tau_{\rm s}}{\tau_{\rm tr}} \right)\, 
\left(\frac{R_c}{a}\right)^2              \label{555}
\end{equation}
the following expressions  
\begin{eqnarray}
\sigma_{xx}^{\rm DP} & \sim & - {\rm sgn}(\Delta \omega) \; 
\pi^2   \left[e^2D_{xx}\nu^* \right]  \frac{\tau^*_{\rm s}}{\tau^*_{\rm tr}}
    \left(\frac{eER_c}{ 4\Delta\omega}\right)^2  \; 
 \sqrt{\frac{\tau_{\rm tr}}{\tau_{\rm s}}}^3 
\left(\frac{a}{R_c}\right)^3        \,, \label{560}  \\
\sigma_{yy}^{\rm DP} & \sim & 
    \left[ e^2D_{yy}\nu^*\right]\left(aB/\pi\sqrt{\tau_{\rm s}^*\tau^*_{\rm tr}}
  \over E_{\rm dc} \right)^2
  \left(eER_c\over 4\Delta \omega\right)^2  
A_2\left(\frac{\Delta \omega}{ 2V_N},
\frac{\Delta \mu}{V_N}\right) \; \sqrt{\frac{\tau_{\rm tr}}{\tau_{\rm s}}} 
\frac{a}{R_c}  \;.
         \label{565}    
\end{eqnarray}
As mentioned below Eq. (\ref{170}), the singularity of $
\sigma_{yy}^{\rm DP}$ is cut off for small {\it dc} fields by
inelastic scattering processes.  This cutoff was discussed extensively
in Ref. \onlinecite{Dietel1}.  By transferring this discussion to the
present case, we obtain a cutoff for small {\it dc } fields at $
E_{\rm dc} \sim E^*_{\rm dc} $ where 
\begin{equation}
 E^*_{\rm dc}=\frac{Ba}{ 2 \pi
\sqrt{\tau_{\rm in} \tilde{\tau}^*_s}} 
\sim  \frac{Ba}{ 2 \pi \sqrt{\tau_{\rm in} \tau_s^*}}  
\left(\frac{\tau_{\rm tr}}{\tau_s}\right)^{1/4}
\left(\frac{a}{R_c}\right)^{1/2} \,.      \label{568} 
\end{equation}
For $ E_{\rm dc} \ll E^*_{\rm dc} $ the photoconductivity crosses over 
to Ohmic behaviour matching with (\ref{565}) for 
$  E_{\rm dc} \sim E^*_{\rm dc} $  
 
In principle we can also get analytical formulas for the scaling laws 
in the transition region 
$ \Delta \omega'_0 <  |\Delta \omega| <\Delta \omega_0 $. Due to their 
complexity we will not give explicit expressions in this range. 
     
\subsubsection{Photoconductivities for the distribution function mechanism}

In the following, we calculate the photoconductivity 
in the regime $ |\Delta \omega| < 
\Delta \omega_0' $.   
By using Eqs. (\ref{176}), (\ref{178}),
(\ref{518}) and (\ref{540}) we obtain with $ (\Delta_{\rm ph}/a)^2 \approx 
(\tau_{\rm s}/\tau_{\rm tr}) (R_c/a)^2 $ that the $ \epsilon $ integral 
(\ref{180}) has its main contribution from the band edges. 
 \begin{eqnarray}
\sigma^{\rm DF}_{xx} & \sim &    4 
  \left(\tau_{\rm in} \over 
 \tau_{\rm tr}^* \right)
 \left[e^2D_{xx}\nu^*\right]
 \left(\frac{eER_c}{ 4\Delta\omega}\right)^2 
  B_1\left(\frac{\Delta \omega}{ 2V_N},
 \frac{\Delta \mu}{V_N}\right) 
 \left(\frac{\tau_{\rm tr}}{\tau_{\rm s}}\right)
 \left(\frac{a}{R_c}\right)^2  \,,
                       \label{570}  \\
   \sigma_{yy}^{\rm DF} & \sim &    4 
\left(\tau_{\rm in}\over\tau_{\rm tr}^*\right)
[e^2 D_{yy} \nu^*] 
    \left({eER_c\over  4\Delta\omega}\right)^2 \, 
 B_2\left(\frac{\Delta \omega}{ 2V_N},
 \frac{\Delta \mu}{V_N}\right)    \frac{\tau_{\rm in}}{\tau^*_{\rm s}}
  \sqrt{\frac{\tau_{\rm tr}}{\tau_{\rm s}}} 
 \frac{a}{R_c}  \;.             \label{575} 
 \end{eqnarray}
For calculating the photoconductivity in $ y$-direction 
$ \sigma_{yy}^{\rm DF} $ we have taken 
into account (\ref{190}), (\ref{540}) and the discussion below (\ref{526}).
This conductivity expression is valid for $ E_{\rm dc} \ll  aB/
\sqrt{\tau_{\rm in} \tilde{\tau}^*_{\rm s}} $ and $ |\Delta \omega |< 
\Delta \omega_0' $. 
By using $ A_2 \sim (\Delta \omega_0'/V_N) $,   
$ B_1 \sim (\Delta \omega_0'/V_N)^{-3/2} $ and  
$ B_2 \sim (\Delta \omega_0'/V_N)^{1/2} $ we obtain for 
$ |\Delta \omega| < \Delta \omega_0' $ with (\ref{555}) 
the scaling relations 
\begin{equation}
  \sigma_{xx}^{\rm DF} \sim    
 \frac{ \tau_{\rm in}}{\tau^*_{\rm s}}
\frac{\tau_{\rm tr}}{\tau_{\rm s}}
\left(\frac{a}{R_c}
\right)^2  
\, \sigma_{xx}^{\rm DP} \quad  \quad , \quad 
\sigma_{yy}^{\rm DF} \sim  \frac{ \tau_{\rm in}}{\tau^*_{\rm s}}
\frac{\tau_{\rm tr}}{\tau_{\rm s}}
\left(\frac{a}{R_c}
\right)^2  
  \sigma_{yy}^{\rm DP}.
\label{580} 
\end{equation}
Summarizing, we obtain that the photoconductivities  in the striped state 
regime are dominated  
by the distribution function mechanism where remarkably the scaling factors 
are the same for both direction. We do not expect a 
strong $ \Delta \omega $ dependence of the exact proportionality factors 
to the photoconductivity expressions above  
for $ |\Delta \omega| \ll 
\Delta \omega_0' $ and for impurity correlation functions which do not 
vary strongly  over $ 1/\xi $ in momentum space. In Fig.~1 we show $ B_1 $, 
$ B_2 $ for the half filled upper Landau level system. We expect from  
the figure zero resistance states for both directions at 
$ \omega \approx \omega_c $. The region of possible zero 
resistance states near $ \omega_c $ is given by $\Delta \omega <0 $ for 
$ \sigma_{xx}^{\rm DF} $ and $\Delta \omega >0 $ for 
$ \sigma_{yy}^{\rm DF} $. 
For the frequency range  $ |\Delta \omega|
\gg \Delta \omega_0 $ we obtain zero photoconductivities.         

\subsection{Calculation of the conductivities in the case 
$ {\bf  \sqrt{\tau_{\rm tr}/\tau_{\rm s}} \, (a/R_c) \lesssim  1} $}
In this subsection we discuss the conductivities in the 
regime $ 1/\xi  \gtrsim a/\ell_B^2 $ . 
In this regime, one has to carry out explicitly the 
summations in the definition of the 
transport times (\ref{510}) and (\ref{530}).  
This can be done with the help of Poisson's summation
formula. 
We do not discuss the results of this calculation in detail.
We obtain conductivity contributions beside the terms discussed in 
section III which have a damping term proportional to $ a \xi/\ell_B^2 $ 
times an  
oscillating function of $ 2 \ell_B^2/a \xi $ with period one. 
This can be understood by the functional dependence of the transport times 
in (\ref{510}) and (\ref{530}) because the additional conductivity expressions 
should be approximately periodic in the  number of 
j-summands. These  are given 
by $ 2 \ell_B^2 / a \xi  $. 

One of the main results of this calculation is that the 
derivatives with respect to $ \Delta \omega $ in Eq. (\ref{150})
and with respect to $ \epsilon $ in Eq. (\ref{195})
(we remind that the derivative has to 
be applied also on  $ \tilde{\tau}_{s}(\epsilon) $ in $ D_{yy}$ (\ref{190})) 
have the effect that the photoconductivities $ \sigma^{\rm DP}_{xx} $ 
and $ \sigma^{\rm DF}_{yy} $ 
get singular for $ N \to \infty $ 
due to the steps in the energy dependent transport times. This is 
because the support  
of $ \tilde{W}(q)  $ is 
zero outside the inverse of the correlation length such that the number 
of $ j $-summands in (\ref{510}) and (\ref{530}) depends strongly on 
the energy $ \epsilon $ of $ \tilde{\tau}(\epsilon)$.      
$ \sigma^{\rm DF}_{yy} $ in Eq. (\ref{195}) 
is singular when
$ \tilde{\nu}^*(\epsilon+\Delta\omega ) $ and the step in 
$  \tilde{\tau}_s(\epsilon) $ is in accordance. 
For  $ \sigma^{\rm DP}_{xx} $ (\ref{160}) we first transform the 
integral in Eq. (\ref{150}) to displacement coordinates corresponding to 
$ \Delta a $ in Eq. (\ref{535}) 
(these were also used in Ref.\ \onlinecite{Dietel1} 
to calculate $ A_1 $). In these coordinates the impurity 
scattering time $ \tilde{\tau}_{2}(\epsilon) $ does not depend 
explicitly on $ \Delta \omega $. Thus we can change the order 
of the derivative with respect to $ \Delta \omega $ and the 
integration in (\ref{150}).  
Keeping this in mind, we find for the positions of the photoconductivity peaks 
\begin{eqnarray}
1-\frac{1}{2}\left(\frac{\Delta \omega}{V_N}\right)^2-
\cos\left(2\pi  \frac{\ell_B^2}{a \xi} \right) & = & 0 
\qquad  \mbox{for} \qquad 
\sigma^{\rm DP}_{xx}         \label{590}  \,,       \\
1-\left(\frac{\Delta \omega}{V_N}\pm 
\cos\left(\pi \frac{\ell_B^2}{a \xi}\right)\right)^2 & = & 0   \qquad  \mbox{for} 
\qquad \sigma^{\rm DF}_{yy} \,.
\label{595}
\end{eqnarray}
It is clear from the discussion above that these singularities 
should not be expected for general filling fraction and $ T \ll V_N $. 
The reason is that the range of the integrals over the 
energy $ \epsilon $ in the photoconductivity formulas, Eqs. (\ref{150}) and 
(\ref{195}), are strongly dependent on the filling fraction. 
Thus, the steps of the energy dependent transport times could  
lie outside this range. Nevertheless, for $ T \gg V_N $ or 
integer filling fraction these singularities should be seen
in general. 
We point out that in real systems the singularities 
of the photoconductivities will be  
smeared out for finite Landau level index $N $ 
due to the smoothness of the Laguerre wavefunctions in the 
monotonic region (see the discussion below Eq. (\ref{a640})).

\section{Electrons in high Landau levels with unidirectional 
periodic modulation of arbitrary strength}

In the last two sections, we calculated the conductivities in $x$- and 
$y$-direction within a theory that uses the eigenvalues and eigenvectors 
calculated in first order perturbation theory in $ \tilde{V} $. 
This is correct for $ V_N \ll \omega_c $. 
In this section we generalize the calculation of the eigenvectors 
and the eigenfunctions to the case of large Landau level index $ N $ 
without any restriction on the relation of $ V_N $ and $ \omega_c $. 
It is well known that one gets the exact eigenvalues and eigenfunctions 
for physical systems within the WKB approximation in the limit of  
large quantum numbers corresponding to high Landau levels 
in our system. 
This is the physical setting  we are interested in.    

In the following, we will calculate in  
subsection A the eigenvalues and 
eigenfunctions of the Hamiltonian of an electron in a homogeneous 
magnetic field and with a 
unidirectional periodic modulation in $x$-direction within the WKB approximation. 
We will use these wavefunctions to calculate simple matrix elements
which are needed for the derivation of the microwave photoconductivities 
\cite{Dietel1}. In subsection B we calculate the matrix elements of 
 plane waves with respect to WKB eigenstates and discuss the implications 
on the calculation 
of impurity averaged absolute square of plane wave matrix elements,
relevant for the calculation of conductivity values.  

\subsection{WKB-approximation for an electron in the background 
of a homogeneous magnetic field and a unidirectional periodic potential}

By carrying out the standard separation ansatz 
$ \langle \vec{r}|n k\rangle=\psi_{nk}(x) e^{i k y}/\sqrt{L_y} $ 
for the eigenfunctions of
the Hamiltonian under study, we obtain that  $ \psi_{nk}(x) $ is an 
eigenfunction of the Hamiltonian 
\begin{equation} 
H=\frac{1}{2m}p_x^2+\frac{1}{2m \ell_B^4} (x-X)^2+\tilde{V}
\cos\left(\frac{2\pi}{a} x\right)   \label{2010} 
\end{equation}   
with $ X=k \ell_B^2 $ and $ m $ is the mass of the electron.  
The eigenfunction of $ H $ corresponding to the eigenvalue $ E_n $ 
is given in the WKB approximation by
\begin{equation} 
\psi_{nk}(x)=\frac{1}{N_{\psi}} \left(\frac{1}{2m(E_{n}-
\tilde{V}_{\rm eff}(x))}\right)^{\frac{1}{4}}
\cos\left(\int\limits_x^{x_R}dx'\sqrt{2m(E_n-\tilde{V}_{\rm eff}(x'))}
-\frac{\pi}{4} \right)            \label{2020}
\end{equation}
with the effective one-dimensional potential 
\begin{equation}
\tilde{V}_{\rm eff}(x):=\frac{1}{2m \ell_B^4} (x-X)^2+
\tilde{V}\cos\left(\frac{2\pi}{a} x\right)   \,.      \label{2030}
\end{equation}
$ N_{\psi} $ is a normalization constant which will be determined below. 
$ x_L $ ($ x_R $) is the  left (right) classical reflection point 
of a particle in the potential $ \tilde{V}_{\rm eff}(x) $ 
with energy $ E_n $, i.e. 
\begin{equation}
 \frac{(x_{L,R}-X)^2}{2m \ell_B^4}+\tilde{V}
\cos\left(\frac{2\pi}{a} x_{L,R}\right)
=E_n \,. \label{2035}
\end{equation}  
The WKB quantization rule is given by:
\begin{equation}
 \int\limits_{x_L}^{x_R}dx'\sqrt{2m (E_n-\tilde{V}_{\rm eff}(x'))}=
\left(n+\frac{1}{2}\right)\pi \,.\label{2040}
\end{equation} 
In the following, we will solve this equation for $ n \to \infty $. 
We will insert a $ \tilde{V} $ expansion 
of the solution for $ x_{L,R} $  in (\ref{2035}),
at the left hand side of (\ref{2040}).
Then it is easily seen that for $ n \to \infty $ it is sufficient 
 to consider   
$ x_{L,R} $ up to the zeros  order of $ \tilde{V} $. This is because the 
correction of $ x_{L,R} $ due to the oscillatory potential $ \tilde{V} $ is
proportional to $ 1/\sqrt{n} $. 
Furthermore, one finds that the integrand of the left hand side of 
(\ref{2040}) has to be considered only up to first order in $ \tilde{V} $ 
for $ n \to \infty$. 
After some calculation, one gets
from Eqs. (\ref{2030}) and (\ref{2040}) for $ n \to \infty $ 
\begin{equation}
E_n=\left(n+\frac{1}{2}\right) \omega_c  + 
\tilde{V} \cos\left(2 \pi \frac{X}{a}\right) 
J_0\left(2 \pi \frac{R_c}{a}\right) \,. \label{2050}
\end{equation}
This energy corresponds to the first order in $ \tilde{V} $   
result (\ref{15}).    

Next, we calculate explicitly the wavefunction $ \psi_{nk} $
(\ref{2020}).  As in the calculation of the spectrum, we will
calculate the integral in the cosine in Eq. (\ref{2020}) by the help
of an expansion of the integrand with respect to $ \tilde{V} $. Then
one can show that it is sufficient to take into account only the
linear order to get all finite terms in the cosine for $ n
\to \infty $.  With the help of an elementary integration, we get
\begin{equation} 
 \psi_{nk}(x)=\frac{1}{N_{\psi}} \left(\frac{1}{2m (E_{n}-
 \tilde{V}_{\rm eff}(x))}\right)^{\frac{1}{4}}
 \sin\left( \Theta^V_{n}(x-X)+C^{V}_{nk}(x-X)+
\frac{\pi}{4}\right)    \label{2060}
\end{equation}
with 
\begin{equation}
 \Theta^V_{n}(x)=(n+1/2) \arccos\left(
\frac{x}{\sqrt{2(n+1/2)}\ell_B}\right) 
 -\frac{1}{2}(n+1/2)\sin\left(2\arccos\left(\frac{x}
{\sqrt{2(n+1/2)}\ell_B}
 \right)
 \right)
 +\frac{\pi}{4}           \label{2065}
\end{equation}
where $ \arccos(x)\in[0,\pi] $.
$ C^{V}_{nk} $ is a term proportional to  $ \tilde{V} $. 
This term is given by 
\begin{equation}
C^{V}_{nk}(x)=-m \ell_B^2 \nts{10} 
\int\limits^1_{x/\sqrt{2(n+1/2)}\,\ell_B} \nts{10}dx'  
\frac{1}{\sqrt{1-x'^2} }\;  \tilde{V} 
\cos\left(\frac{2 \pi}{a} \left(\sqrt{2(n+1/2)}\,\ell_B \, x'+X\right)\right) \,.
\label{2070}
\end{equation}
By the help of a Fourier expansion we find
\begin{align}
&  C^{V}_{nk}(x)=m \ell_B^2 \tilde{V} 
\bigg\{-J_0\left(2\pi \frac{R_c}{a}\right) 
\cos\left(2 \pi \frac{X}{a}\right)
\left(\frac{\pi}{2}- \varphi_{x} \right) 
\label{2075}  \\
& +\sum\limits_{l >0 }\frac{2}{2l-1} J_{2l-1}\left(2\pi \frac{R_c}{a}\right) 
\sin\left(2 \pi \frac{X}{a}\right)
\cos\left((2l-1)\varphi_{x}\right)  \nonumber \\
& 
\qquad +\frac{2}{2l} J_{2l}\left(2\pi \frac{R_c}{a}\right) 
\cos\left(2 \pi \frac{X}{a}\right)
\sin\left(2l \varphi_{x} \right) \bigg\}   \nonumber 
\end{align} 
with $ \varphi_x=\arcsin(x/\sqrt{2(n+1/2)} \ell_B) $.
It should be emphasized that $ C^V_{nk} $ is a finite  term 
for $ n  \to \infty $. 

Next, we calculate the normalization constant $ N_{\psi} $.  This
constant is determined by the requirement that $ \psi_{nk}(x) $ has
norm one. Thus, we calculate the one-dimensional integral of the
function $ \psi^2_{nk}(x) $ for $ n \to \infty $. 
One can neglect in $ \psi^2_{nk}(x) $ the
$\tilde{V} $ term in the prefactor $ 1/(E_n-\tilde{V}_{\rm
  eff}(x))^{1/2} $ of $ \psi^2_{nk}(x) $ during this calculation.
Furthermore, the integrand contains a factor $ \sin^2[\ldots] $ whose
argument is given by the argument of the sine in Eq. (\ref{2060}).
For large Landau levels, ($ n \to \infty $), we have $
\sin^2[\ldots]=(1-\cos[\ldots])/2 \approx 1/2 $. This replacement is
exact in leading order in $n$ of $ N_{\psi} $ for $ n \to \infty $,
since $ \cos[\ldots] $ is a fast oscillating function for $ n \to
\infty$.  After carrying out the integral, we obtain
\begin{equation}
N_{\Psi}=\sqrt{\frac{\pi}{2}} \ell_B \,.         \label{2080}
\end{equation}
Now, we can compare $ <\vec{r}|nk>=\psi_{nk}(x) e^{i k y}/\sqrt{L_y} $   
for $ \tilde{V}=0 $, i.e. $ C_{nk}^{V}(x)=0 $, with the exact 
eigenfunctions of the Hamiltonian in the Landau gauge. As is well known, 
these eigenfunctions consist of Hermite polynomials. 
By using the identities 
\cite{Erdelyi} $ H_{2m}(x)=(-1)^m 2^{2m} m! L_m^{-1/2}(x^2) $, 
$ H_{2m+1}(x)=(-1)^m 2^{2m+1} m! L_m^{1/2}(x^2) $ and further (\ref{a630}) 
and (\ref{a640})  
for the asymptotics of the Laguerre polynomials we get accordance of the 
WKB wavefunctions and the exact eigenfunctions for 
Landau level $n \to \infty$ and $ \tilde{V}=0 $. 

Next we calculate the matrix element 
$ \langle n k|x|n'k'\rangle   $. 
For calculating this quantity we have to determine 
$ \Theta^V_{n+\Delta n k} $ for 
a fixed $ \Delta n $ and $ n \to \infty $.  
With the help of a Taylor expansion we get 
\begin{equation}
\Theta^V_{n+\Delta_n}\approx \Theta^V_{n}+
\Delta n \arccos\left(\frac{x}{\sqrt{2(n+1/2)}\, \ell_B} \right) \,.  
\label{2090}
\end{equation} 
With the integration techniques stated below Eq. (\ref{2075}), we 
get for $ n \to \infty $ 
\begin{equation} 
 \langle n k|x|n' k'\rangle =\frac{R_c}{2} \; \delta_{k,k'} \;
 \left(\delta_{n,n'+1}+\delta_{n,n'-1} \right) \,.   \label{2100} 
 \end{equation} 
This corresponds exactly to the first order perturbation result 
in $ \tilde{V} $ \cite{Dietel1}. 
We should mention that one can derive similarly the expected 
orthonormal basis property of the $ |n k\rangle  $ eigenstates for $ n \to \infty $.  \\
\hspace{0.2cm}

As was shown in  Ref. \onlinecite{Dietel1} as well as in section III,
when calculating the various dark and photoconductivities 
one has to know the eigenstates of the system with the unidirectional 
periodic potential (considered above) and an additional {\it dc} field in 
$x$- or $y$-direction. For calculating the linear response on this additional 
{\it dc} potential it is enough to consider the first order correction of the 
wavefunction and the energy due to this field. 
The generalization of the WKB approximation for the system 
with an additional
{\it dc} field in $x$-direction is straightforward. It amounts to trivial change in  wavefunctions and energies  
as in the well known transformations for the  system without 
the applied periodic modulation potential. Next we calculate 
the eigenfunctions and 
eigenvalues in the case of the {\it dc} field in the $y$-direction. 
This was done 
in Ref. \onlinecite{Dietel1} for $ V_n \ll \omega_c $. One can carry 
out a similar calculation in the case of Landau level index $ n \to \infty $
without restrictions on $ V_n $ and $ \omega_c $  
by using for the Landau level basis set named $ |nk> $ in 
Ref. \onlinecite{Dietel1} 
the WKB basis set calculated above (see for example 
(\ref{a10}) for the eigenfunctions).  
  
Finally, we want to give a short  application of our findings 
to the striped state physics. The striped state was experimentally identified 
by the measurement of an anisotropy in the resistivity of 
quantum Hall systems 
in high Landau levels $ N \ge 3 $.\cite{Lilly,Du2} This anisotropy 
disappears above Landau levels $ N \approx 7 $. The disappearance of 
the striped state at very high Landau levels can be understood by 
considering the spectrum, Eq. (\ref{2050}).  
By taking into account that, at thermodynamic equilibrium, 
all states are occupied with energy values lower than the chemical potential 
and further that the period of the striped state is $ a \approx 2.6 R_c $ 
we obtain that the striped state starts to diminish 
at $ V_N \approx \omega_c $. 
By using  $ \omega_c = 2 \pi e \rho/ N m $ where $ \rho $ is the 
density of the electron system and further 
due to the Poisson formula 
of electrodynamics 
$ V_N \sim 
\tilde{V} \approx a 
e^2 \rho/ N \approx  2.6 \, e^2 R_c \rho/N $ where 
by approximation the electron density in the highest occupied Landau level 
is given by $ (\rho/ N) $, we obtain a vanishing of the striped 
state at high Landau levels. In order to determine the correct Landau level index 
$ N $ of the transition, one should take into account the screening
of the completely filled lower Landau levels as well as the total 
Hartree-Fock potential of the upper Landau level \cite{Fogler}. This trespasses the scope of this work.      
    
\subsection{Matrix elements of plane waves with respect 
to WKB wavefunctions and  implications on the 
calculation of impurity averaged transition rates} 

In this subsection, we give a short outline of the calculation of 
matrix elements of plane waves with respect to the WKB states $ |n k\rangle  $.
These matrix elements are calculated explicitly in App. D. 
As is well known, the 
matrix elements of plane waves with respect to the Hermite polynomials 
which are the exact eigenstates for $\tilde{V}=0 $ are given by Laguerre 
polynomials. As is outlined in App. D, we obtain the asymptotic forms 
given in App. C for these matrix elements when using the WKB approximated 
wavefunctions for $  \tilde{V}=0 $. 
These asymptotic 
forms of the Laguerre polynomials, for large Landau level index $n $, consist
 in the relevant oscillatory region (see App.~C) 
of a rapidly oscillating factor $ \sin(\Theta_n)  $ (Eq. (\ref{a630})) 
times a non-oscillatory amplitude. Here, $ \Theta_n(x) $ is defined as in Eq. 
(\ref{a630}) 
where the argument $ x $ is proportional to the absolute square
of the wavevector of the plane wave. 
When switching on $ \tilde{V} $ we obtain for the WKB matrix elements 
a separation of the $ \sin(\Theta_n)  $ asymptotics in two phase factors,  in the following form 
\begin{equation}     
\sin(\Theta_n(x)) \rightarrow \exp[i \Theta_n(x)]  
\exp[\,i {\cal V}^+_n(x)] + \exp[\,-i \Theta_n(x)] 
\exp[\,-i {\cal V}^-_n(x)]            \label{2120}
\end{equation} 
For large Landau 
level index $n$, the functions $ \exp[\pm i{\cal V}^\pm_n(x)] $ are slowly oscillating functions in $x$.
The exact result is given  in Eq. (\ref{a1030}). 
In contrast to the calculation of the matrix element of the 
$x$-operator (\ref{2100}) and the eigenenergies (\ref{2050}), we thus obtain a 
$ \tilde{V} $-dependence of the plane wave matrix elements.
 
In the following, we will provide arguments that, despite this 
$ \tilde{V} $-dependence of the plane wave matrix elements, we do not 
have a $ \tilde{V} $-dependence of the impurity averaged 
square of these matrix elements 
relevant within the calculation of the conductivity \cite{Dietel1}. 
The impurity averaged square of the plane wave matrix elements enters the conductivity  expressions via the various scattering times 
\cite{Dietel1}.  
We note  
that one gets the exact transport 
scattering times in high Landau levels 
when one uses the asymptotics (\ref{a630}) and (\ref{a640})
for the Laguerre polynomials. 
This was shown for smooth disorder in Ref. \onlinecite{Dietel1} but can 
be generalized to the case without restrictions on the correlation length 
$ \xi $. The calculation is similar to 
the calculation of the normalization constant of WKB states 
below (\ref{2070}) due to the resembling of 
Laguerre and Hermite polynomial asymptotics.   
From this we obtain with (\ref{2120}) that the impurity 
averaged square of the plane wave matrix elements remains the same 
for large Landau level index $n $,
irrespective of the value of $ \tilde{V} $ when using 
the WKB matrix elements. 

In summary, we obtain that all inputs in the conductivity calculations,
i.e. the eigenenergies (\ref{2050}), the $x$-operator 
matrix element (\ref{2090}) as well the various transport time integrals, 
do not depend on the oscillatory strength $ \tilde{V} $ for large Landau level 
index $ n $. Thus, we can easily generalize the 
calculations carried out in Ref. \onlinecite{Dietel1} and in the last 
sections for the various 
conductivities to general $ V_N $ and $ \omega_c $. 
It will become clear soon that the problem 
of calculating the various conductivities then is a mere summation problem. 
In the limit $ V_N \gg \omega_c $ we can carry out these 
summations easily.

\section{Conductivities for {\boldmath $ V_N \gg \omega_c $}}
We calculated in section III and IV the conductivities in $x$- and
$y$-direction within a theory that uses the eigenvalues and
eigenvectors in first order perturbation theory in $ \tilde{V} $.
This is correct in the limit $ V_N \ll \omega_c $. In this section, we
will calculate the various conductivities in the opposite limit, $ \mu
\gg V_N \gg \omega_c $, which excludes a perturbative treatment in
$V_N$. Due to the results of the last section we can still use 
the formalism derived in Ref. \onlinecite{Dietel1} for calculating the 
various conductivities but one has to take into account the 
overlapping of the Landau levels in the calculation of the 
transition probabilities. 

It is clear, that the experimental situation where $ V_N \ll \omega_c $ 
considered up to now corresponds to the situation of well separated 
Landau levels in the microwave experiments without periodic modulation. 
This was also seen in the same scaling of the conductivity expressions 
of both systems  \cite{Dietel1}.
As derived in the last section, the effective Landau level width  
in our system is proportional 
to $ V_N \sim \tilde{V} B a  / \sqrt{\rho} $. 
In Weiss oscillation experiments, it is 
possible to vary $ \tilde{V} $ by changing the grid potential
\cite{Weiss}.
Thus, we obtain that by changing $  \tilde{V} $ we can 
reach the situation of large Landau level overlapping which is 
the low magnetic field regime  in the microwave experiments without 
periodic modulation.  
The regime $ V_N \gg \omega_c $ is not relevant for the striped state, 
so that we can restrict ourselves to the case $ a \ll R_c, \ell_B^2/\xi  $.
Furthermore, we restrict us to the temperature regime 
$ \mu \gg T \gtrsim V_N \gg \omega_c $ which is  the typical experimental 
regime for the system without periodic modulation.   
In the following, we calculate the dark 
conductivities in subsection A and 
the photoconductivities in subsection B. 
The density of states, Eq. (\ref{20}),  
and the dimensionless density of states, Eq. (\ref{45}),  
for $ V_N \gg \omega_c $ are denoted by $  \nu^*_{\Sigma}(\epsilon) $ and  
$\tilde{\nu}^*_{\Sigma}(\epsilon) $, respectively. 

We obtain for the density of states 
\begin{equation} 
\nu_{\Sigma}^*(\epsilon)= 
 \sum_{n \in \mathbb{Z}} 
\nu^*_{N}(\epsilon+n \omega_c)
\approx 
\nu_0 + 2 \nu_0 \sum_{n>0} 
J_0\left(\frac{2 \pi n V_N}{\omega_c}\right) 
\cos\left(
 \frac{2 \pi n \epsilon}{\omega_c}\right) + 
O\left(\frac{\omega_c}{V_N}\right)\,. \label{2530} 
\end{equation}  

By using the Bessel asymptotic expansion 
$ J_0(x)\approx \sqrt{(2 /\pi x)} \cos(x-\pi/4) $ for large $x $ 
we obtain 
\begin{eqnarray} 
\nu_{\Sigma}^*(\epsilon) & \approx  
& \nu_0 + 2 \nu_0
\sum_{n>0} 
\left(\frac{\omega_c}{n \pi^2 V_N}\right)^{1/2} 
\cos\left(\frac{2 \pi n V_N}{\omega_c} 
-\frac{\pi}{4}\right) \cos 
\left(\frac{2 \pi n\epsilon}{\omega_c} \right)+
O\left(\frac{\omega_c}{V_N}\right)  \nonumber \\
&= & \nu_0+ \nu_0 \; \left(\frac{ \omega_c}{2 \pi^2 V_N}\right)^{1/2}
\, \Sigma\left(\frac{V_N}{\omega_c} 
,\frac{\epsilon-\omega_c/2}{\omega_c}\right)+O\left(\frac{\omega_c}{V_N} 
  \right)              \label{2532}
\end{eqnarray}
where the oscillatory part of the dimensionless density 
of states is given by 
\begin{equation} 
\Sigma(x,y)=   
\sum_{\sigma \in \{\pm\}} 
\zeta\left[\frac{1}{2}, \left(x+\sigma y\right)\rm{mod} \, 
  1 \right]  \label{2535}
\end{equation} 
Here, $ \zeta $ is the Riemann zeta function. 
The $ \sqrt{\omega_c/V_N} $ correction to the 
density of states is oscillatory in $ \epsilon $ 
with a wavelength $ \omega_c $ and has singularities for 
$  (V_N \pm (\epsilon-\omega_c/2))/\omega_c \;{\rm  mod } \;  1 =0 $.
These  energy values correspond to the edges of the 
Landau band where the density of state contribution is singular (\ref{40}).
In realistic systems these singularities are cut at 
$ \Delta \epsilon  \sim \sqrt{\omega_c/\tau_s} $ due to impurity scattering 
when $  \omega_c>\sqrt{\omega_c /\tau_s}  $.    
The second term in (\ref{2535}) is then further changed by an additional  
damping factor $ \sim \sqrt{\omega_c \tau_s } $ in the case 
$ V_N \gg   \sqrt{\omega_c /\tau_s} > \omega_c  $ by taking into account 
the Landau level broadening $  \sqrt{\omega_c /\tau_s}  $ for the 
system without modulation.

 \begin{figure}
 \begin{center}
\psfrag{Sigma}{\scriptsize{$\Sigma$}}
\psfrag{A11111111111111111111111}{\scriptsize $(V_N/\omega_c)_1$=0.}
\psfrag{A22222222222222222222222}{\scriptsize $(V_N/\omega_c)_1$=0.25,0.75}
\psfrag{A33333333333333333333333}{\scriptsize $(V_N/\omega_c)_1$=0.5}
 \psfrag{epsilon/omegac}{\hspace{-0.3cm} 
\scriptsize{$(\epsilon/\omega_c-1/2)_1$}}
 \includegraphics[width=8cm]{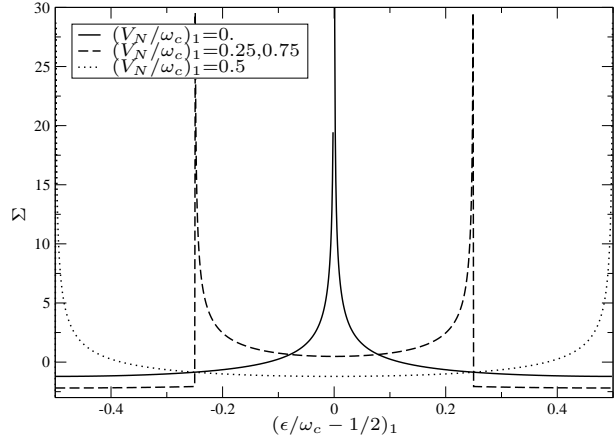}
 \end{center}
\vspace{-0.2cm}
 \caption{The oscillatory part of the 
density of states 
$ \Sigma(V_N/\omega_c,\epsilon/\omega_c-1/2) $ (\ref{2535})  
where $ (V_N/\omega_c)_1:=(V_N/\omega_c)\, {\rm mod }\,1 $.  
The values where $ (\epsilon/\omega_c)_1-1/2 =0 $ correspond to the center 
of the Landau bands.}  
 \end{figure}
Fig.~2 depicts this oscillatory correction to the density of states 
leading  to oscillatory contributions in the conductivities. 
We point out the similarity of the density of states formula (\ref{2532}) 
with the corresponding formula of the system 
without periodic modulation in the overlapping Landau level regime. 
In this system the density of states  is given by 
$ \nu=\nu_0-2 \nu_0 \delta \cos(2 \pi \epsilon/\omega_c)$ with a 
damping term $ \delta = \exp[-\pi/\omega_c \tau_s]$. When comparing 
this expression with the density of states formula (\ref{2532}) it seems  
suggestive that we could get the conductivities in the direction 
of the periodic modulation up to numbers from 
the system without periodic modulation when carrying out the 
the replacement $ \delta \rightarrow \sqrt{\omega_c/V_N} $ in the 
corresponding formulas. That this is in fact true will be shown below.
    
\subsection{Calculation of  dark conductivities}

We start with a  discussion of the dark conductivity.
With Ref. \onlinecite{Dietel1}, 
we obtain for the conductivities per energy  
$ \sigma_{xx} (\epsilon)$ and  $\sigma_{yy}(\epsilon)$, Eq. (\ref{110}),
\begin{equation} 
\sigma_{xx}(\epsilon)=e^2 D_{xx} \nu^* \, (\tilde{\nu}_{\Sigma}^*(\epsilon))^2
\approx e^2 D^{\Sigma}_{xx}
\nu_0\left(1+ 
 \left(\frac{2 \omega_c}{\pi ^2 V_N} \right)^{1/2} \; 
\Sigma\left(\frac{ V_N}{\omega_c}, \frac{\epsilon-\omega_c/2}{\omega_c}\right)
\right)   \,,
 \label{2540}
\end{equation}  
\begin{equation}
\sigma_{yy}(\epsilon)=   e^2 
  D_{yy} \nu^* \frac{1}{\tilde{\nu}_{\Sigma}^*(\epsilon)}
\sum_{n \in \mathbb{Z}} 
\frac{1}{\tilde{\nu}^*_{N}(\epsilon +n \omega_c) }    
  \approx    e^2    D^{\Sigma}_{yy} \nu_0  \left(1 -  
\left(\frac{ \omega_c}{2 \pi^2 V_N}\right)^{1/2} 
\Sigma\left(\frac{ V_N}{\omega_c}, 
\frac{\epsilon-\omega_c/2}{\omega_c}\right)\right) 
.     \label{2560}
\end{equation}
where we used 
$ \sum_{n \in \mathbb{Z}} 
1/\tilde{\nu}^*_{N}(\epsilon +n \omega_c)  =  
\pi V_N/2 \omega_c +O(1) $.
Note that $ D_{xx} $, $ D_{yy} $ are defined by Eq. (\ref{103}) for $
\epsilon =E_n $. The diffusion constants $ D^{\Sigma}_{xx} $, $ D^{\Sigma}_{yy} $
for $ V_N \gg \omega_c $ are given by  
\begin{equation} 
D^{\Sigma}_{xx}  =  
\frac{R_c^2}{2\tau_{\rm tr}} \quad \mbox{and} \quad 
D^{\Sigma}_{yy}= 
 \left\langle \frac{1}{(\pi a \nu^*)^2} \right\rangle  
\tau_s = \frac{1}{2} \frac{1}{(\pi a \nu^*)^2} \,  
\tau_s                  \,.  \label{2565}
\end{equation}
That $ D^{\Sigma}_{xx} $ and $ D^{\Sigma}_{yy} $ are in fact the diffusion 
constants for $ V_N \gg \omega_c $ can be shown 
by taking into account Eq. (\ref{103}) 
with the replacements that the scatter time $ \tau_s^* $ 
should be substituted by $ \tau_s^* \nu^*/ \nu_\Sigma^* \approx \tau_s$ in 
these expressions and 
furthermore the expression for  $ D_{yy} $ in (\ref{103}) should 
be averaged over the various Landau levels intersecting one energy level
weighted by the corresponding density of states contribution. 
We point out that the second term in $ \sigma_{xx}(\epsilon) $ (\ref{2540})
corresponds to the 
Shubnikov de-Haas term of the conductivity without periodic modulation. 

\subsection{Calculation of photoconductivities}
In this subsection, we calculate the various photoconductivities for $
\mu \gg T \gtrsim  V_N \gg \omega_c $ by using the methods presented in Ref.
\onlinecite{Dietel1}.

\subsubsection{Photoconductivities for the displacement mechanism}

The photoconductivity $ \sigma_{xx}^{\rm DP}$ 
for $ V_N \gg \omega_c $ is given by (\ref{140}) with 
\begin{equation}
    A_1=  \frac{6}{\pi^2}
\int\limits_0^{\infty} d\epsilon \, \left[f(\epsilon)-f(\epsilon+\omega)
\right]  \; \frac{\partial}{\partial \omega}  
\tilde{\nu}^*_{\Sigma}(\epsilon) 
\tilde{\nu}^*_{\Sigma}(\epsilon +\omega)  \label{2580}
\end{equation}  
Here we used that the impurity averaged matrix element 
$ \int d^2q \, (q^2_y\ell_b^4)\,
 [L^m_{N+1}((q\ell_b)^2/2)-L^m_{N}((q\ell_b)^2/2)]^2 \tilde{W}(q)$ 
does not depend on $m $ which can be shown easily by using the asymptotic 
form (\ref{a640}) for the Laguerre polynomials or by an explicit calculation 
of the integral in App. B for delta correlated impurities resulting in 
$ 12 \pi N$. 
Now we insert the density of states  
expression (\ref{2530}) in (\ref{2580}). By 
taking into account that 
$ \int d\epsilon  \cos(2 \pi n \epsilon/\omega_c) 
(\partial/\partial \epsilon)  f(\epsilon) \sim \exp[-2 \pi^2 n 
T/\omega_c] $ is 
negligible for $ T \gg \omega_c $ we obtain only non-exponentially 
vanishing corrections of the form 
$ \int d\epsilon  \cos^2(2 \pi n \epsilon/\omega_c) 
(\partial/\partial \epsilon) f(\epsilon) \approx -1/2 $. 
Then by taking the cos asymptotics for the Bessel 
function in (\ref{2530}) one can 
carry out the resulting Bessel series giving 
\begin{equation}
  \sigma_{xx}^{\rm DP} \backsimeq   -\frac{12}{\pi}
 \left[e^2 D^{\Sigma}_{xx}\nu_0 \right] 
{\tau_{\rm s}\over\tau_{\rm tr}}  
    \left(\frac{eER_c}{ 4\Delta\omega}\right)^2 
\left(\frac{\omega_c}{V_N}\right)  \frac{\omega}{\omega_c}
\frac{\sin \left(\frac{2 \pi \omega}{\omega_c}\right)}{1-\cos\left(\frac{2 \pi \omega}{\omega_c}\right)} \;. 
   \label{2590}
\end{equation}
 
Next, we calculate the displacement photoconductivity 
$ \sigma_{yy}^{\rm DP}$ keeping in mind 
the remarks below Eq. (\ref{2100}) concerning the WKB wave functions 
with a {\it dc} field in the $y$-direction. 
With Eq. (\ref{160}) we find  
\begin{equation}
   A_2 =\sum_{m,n \in \mathbb Z} 
\frac{m \omega_c-\omega}{V_N^2} 
\int\limits_{0}^{\infty}  \, d\epsilon \; \tilde{\nu}_N^*(\epsilon) 
\tilde{\nu}_N^*(\epsilon+m \omega_c-\omega) 
[f^y_{nl}-f^y_{(n+m)\,l}] \,. \label{2600}
\end{equation}
By taking into account $ \sum_{n=0}^{\infty} [f^y_{nl}-f^y_{(n+m)\,l}]=m $ 
one can easily carry out the $m $-sum for $ V_N/\omega_c \gg 1  $
in Eq. (\ref{2600}) after having done the   
$ \epsilon $ integration. This results in
\begin{equation}
  \sigma_{yy}^{\rm DP} \approx  \frac{(16+32 G)}{\pi^2}  
 \left[ e^2 D^{\Sigma}_{yy}\nu_0\right]
\left(aB/\pi\sqrt{\tau_{\rm tr}\tau_{\rm s}}
  \over E_{\rm dc} \right)^2
  \left(eER_c\over 4\Delta \omega\right)^2 \,.              \label{2610}
\end{equation} 
where $ G \approx 0.915$ is Catalan's constant. 
As discussed in Ref. \onlinecite{Dietel1}, Eq. (\ref{2610}) is only valid 
in the case that the electric {\it dc} field $E_{\rm dc} $ is not too small. 
The reason lies in inelastic scattering which we do not take into 
account when calculating Eq. (\ref{2610}). By carrying out a similar analysis 
as in Ref.\ \onlinecite{Dietel1}, we find that Eq. (\ref{2610}) 
is valid only for 
\begin{equation}  
 E_{\rm dc} \gg E^*_{\rm dc} = 
\frac{B a}{2 \pi \sqrt{\tau_{\rm in}\tau_s  }} \,.   \label{2615}  
\end{equation} 
In the case $ E_{\rm dc} \ll E^*_{\rm dc} $ we have to substitute  
$ E_{\rm dc} $ in the denominator in (\ref{2610}) by  
$ E^*_{\rm dc} $.  
Finally, we remark that the results, Eqs. (\ref{2610}) and (\ref{2615}),
are correct generally for $ T \ll \mu $, since  
$ f^y_{nl} $ is independent of $ l $ for fixed $n $.    
 
\subsubsection{Photoconductivities for the distribution function mechanism}
 
For calculating $ \sigma_{xx}^{\rm DF} $ as well as 
$ \sigma_{yy}^{\rm DF} $ one has to calculate 
the change in the distribution function 
 due to absorption of microwaves. If one assumes that there is no heating,  
 relaxation processes have to be taken into account.  
In the case $ V_N \ll \omega_c $, we calculated the  distribution function 
(see Ref. \onlinecite{Dietel1}) within the relaxation time approximation. 
In the following, we carry out a similar calculation for the 
distribution function in the case $ V_N \gg \omega_c $.
By taking into account the overlap of Landau levels due to the 
large amplitude of the unidirectional modulation potential  
 we get 
\begin{equation}
\label{2620}
        \delta f(\epsilon) \approx  2 \frac{\tau_{\rm in}}{\tau_{\rm tr}}
 \left(\frac{eER_c}{ 4\Delta\omega}\right)^2
\sum_{\sigma=\pm}[ f(\epsilon-\sigma \omega) - f(\epsilon)] \;  
\frac{\nu^*_{\Sigma}(\epsilon-\sigma\omega)}{\nu_0} \,.
\end{equation} 
By using this distribution function, we obtain
for the photoconductivity $ \sigma_{xx}^{\rm DF} $ (\ref{178}) with 
 \begin{equation}
 B_1= - \frac{1}{2} \sum_{\sigma \in \{\pm\}}
\int\limits_{0}^{\infty}
d \epsilon \; 
(\tilde{\nu}_{\Sigma}^*(\epsilon))^2 
\frac{\partial}{\partial \epsilon} \left(
\tilde{\nu}_{\Sigma}^*(\epsilon-\sigma \omega)
[f(\epsilon  -\sigma \omega)-f(\epsilon)] \right)\,. 
\label{2625}
\end{equation}  
The calculation of this expression can be done similar to the calculation 
of $ \sigma_{xx}^{\rm DP} $ above. This results in
\begin{equation}
\sigma^{\rm DF}_{xx} \approx -\frac{16}{\pi} 
 \left(\tau_{\rm in} \over 
\tau_{\rm tr} \right)\left(\frac{eER_c}{ 4\Delta\omega}\right)^2 
\left[e^2 D^{\Sigma}_{xx}\nu_0\right]  
\left(\frac{\omega_c}{V_N}\right) 
\frac{\omega}{\omega_c}
\frac{\sin \left(\frac{2 \pi \omega}{\omega_c}\right)}
{1-\cos\left(\frac{2 \pi \omega}{\omega_c}\right)} \,. 
          \label{2635}
\end{equation}

Finally, we calculate the photoconductivity $ \sigma_{yy}^{\rm DF} $
which is given by Eq. (\ref{190}) with 
\begin{equation}
   B_2 =
\frac{1}{2} \sum_{\sigma \in \{\pm\}}
\int\limits_{0}^{\infty}
d \epsilon \; \sum\limits_{n \in \mathbb{Z}} 
 \tilde{\nu}_{\Sigma}^*(\epsilon-\sigma \omega)   
 [f(\epsilon  -\sigma \omega)-f(\epsilon)]  
\; 
 \frac{\partial}{\partial \epsilon}  \left(
 \frac{1}{\tilde{\nu}_{\Sigma}^*(\epsilon)
 \tilde{\nu}_{N}^*(\epsilon+n \omega_c) }   
 \right)
\label{2640}
\end{equation}
resulting in 
\begin{equation}
\label{2660}
    \sigma_{yy}^{\rm DF}\approx 
-\frac{8}{\pi}  \left(\tau_{\rm in}\over\tau_{\rm tr}\right)
    \left({eER_c\over  4\Delta\omega}\right)^2[e^2 D^{\Sigma}_{yy} \nu_0]\,   
\left(\frac{\omega_c}{V_N}\right) 
\frac{\omega}{\omega_c}
 \;  \frac{ \sin \left(\frac{2 \pi \omega}{\omega_c}\right)}{1-\cos\left(\frac{2 \pi \omega}{\omega_c}\right)}  
\end{equation}
for large $ V_N /\omega_c $. 

By taking into account the above expressions for the photoconductivities
we obtain the result that the photoconductivities 
$ \sigma_{xx}^{\rm DP} $, $ \sigma_{xx}^{\rm DF} $ and 
$ \sigma_{yy}^{\rm DF} $ are oscillating  in $ \omega/\omega_c $ 
whereas the photoconductivity 
$ \sigma_{yy}^{\rm DF} $ is non-oscillating with positive values. 
We obtain the following scaling relations for the photoconductivities in the 
regime $ V_N \gg \omega_c $  
\begin{equation}
\sigma_{xx}^{\rm DF} \sim \,  \frac{\tau_{\rm in}} {\tau_{\rm s}} 
\sigma_{xx}^{\rm DP} \quad , \quad \sigma_{yy}^{\rm DF}
\sim  \frac{\omega}{V_N}\,
\sigma_{yy}^{\rm DP}    \,.        \label{2665}         
\end{equation} 
Thus we find that the photoconductivity 
parallel to the modulation direction 
is dominated by the distribution function contribution. 
In the direction perpendicular 
to the modulation, the photoconductivity is dominated by the 
displacement contribution in the regime $ V_N \gg  \omega$ 
being always positive, 
or oscillating which should result also in zero resistance states   
in the regime $ V_N \ll \omega $. The careful reader could 
have the objection  
that our derivations so far were done for 
$ \omega \approx \omega_c $. We will show in the next section that the 
photoconductivites calculated in this section are also valid for all 
$ \omega > 0 $ up to a non-oscillating positive prefactor which has 
the same scaling 
for all photoconductivities $ \sigma^{\rm photo} $ 
depending on the polarization and 
$ \omega, \omega_c $. 
The frequency dependent scaling of the photoconductivities is then given by 
\begin{equation} 
 \sigma_{xx}^{\rm DP,DF}, 
 \sigma_{yy}^{\rm DF} \sim  C
=-\frac{\omega_c}{\omega}
\frac{\left(1- \frac{\omega}{\omega_c} \right)^{-2}
 \sin \left(\frac{2 \pi\omega}{\omega_c}\right)} 
 {\left(1-\cos\left(\frac{2\pi \omega}{\omega_c}\right)\right)} 
 \quad, \quad \
\sigma_{yy}^{\rm DP} \sim D
 =\left(\frac{\omega_c}{\omega}\right)^2
 \left(1-\frac{\omega}{\omega_c}\right)^{-2} 
\label{2668}
\end{equation}
Here we take 
into account the additional factor $ (\omega_c/\omega)^2 $ for the 
photoconductivities derived in 
section VII below such that 
the scaling functions are valid for all $ \omega >0 $.
In Fig.~3 we show the scaling functions $ C(\omega/\omega_c) $ and 
$ D(\omega/\omega_c) $.  

\begin{figure}
 \begin{center}
 \psfrag{y}{}
 \psfrag{x} 
 {\scriptsize{\hspace*{-0.5cm} $\omega/\omega_c$}}
 \includegraphics[width=8cm]{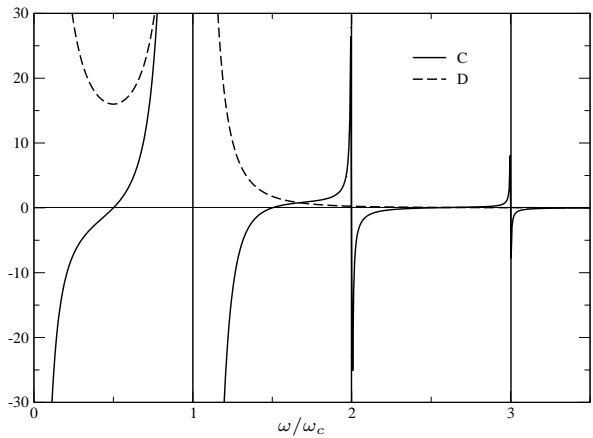}
 \end{center}
 \caption{The frequency dependent scaling $ \sigma_{xx}^{\rm DP,DF}, 
\sigma_{yy}^{\rm DF} \sim  C(\omega/\omega_c) $  (solid line) and 
of $ \sigma_{yy}^{\rm DP} \sim   D(\omega/\omega_c) $ (dashed line). 
The scaling functions
$ C$ and $ D$ (\ref{2668}) are the exact frequency dependent prefactors of the 
photoconductivities  in the case of a circular polarized microwave field 
derived in section VII ($ \alpha=\pi/4 $ and $ \beta=-\pi/4 $).
 }  
 \end{figure}

Next, we compare $ \sigma_{xx}^{\rm DP}, \sigma_{xx}^{\rm DF} $ 
calculated above with the corresponding terms of Ref. \onlinecite{Vavilov1}
for $  \sigma_{xx}^{\rm DP} $ and Ref. \onlinecite{Dmitriev1} for $  
\sigma_{xx}^{\rm DF} $ representing the photoconductivities for the system 
with no modulation potential 
by using the damping factor substitution 
$ \delta \rightarrow \sqrt{\omega_c/V_N} $. We get immediately 
correspondence of the expressions up to numbers as was also the case 
in the non-overlapping Landau level regime shown in Ref. 
\onlinecite{Dietel1}. Nevertheless the form of the photoconductivities 
differ. Especially the singularities of $ \sigma_{xx }^{\rm DP} $, 
$ \sigma_{xx}^{\rm DF} $ and 
$ \sigma_{yy}^{\rm DF} $ at  $ \omega = \mathbb{Z}\omega_c  $ (see Fig.~3)
having their reason in the singularities of the  
density of states (\ref{2532}). They  
are not existent in the system without modulation.
 From the discussion 
below (\ref{2535}) and the calculation of the photoconductivities above 
we obtain that for $ \omega_c > \sqrt{\omega_c/\tau_s} $ the 
photoconducitites are cut at $ |\omega-\mathbb{Z}\, \omega_c| \sim 
\sqrt{\omega_c/\tau_s} $.  For 
 $ V_N \gg \sqrt{\omega_c/\tau_s} > \omega_c $ the   
 singularities in photoconductivities are smeared out completely 
where the conductivity
expressions obtain an additional damping factor 
$ \sim (\omega_c \tau_s) $ due to the damping of the oscillatory part 
of the density of states. As mentioned below (\ref{200}) the 
$ (e E R_c/4 \Delta \omega)^2 $ prefactor in the 
photoconductivity expressions should  be cut at $|\Delta \omega| 
\sim 1/\tau_{\rm tr} $.

\section{Photoconductivities for arbitrary polarization and frequency}
In this section, we generalize our results
to the case of microwave irradiation of frequency $ \omega \gtrsim \omega_c $ 
and general  polarization. An arbitrarily polarized microwave 
field is of the form 
\begin{equation}
\vec{E}(\alpha,\beta) = 
E \left(\cos(\alpha) \cos(\omega t-\beta ) \vec{e}_x 
+\sin(\alpha) \cos(\omega t+\beta) \vec{e}_y \right)  \,.  \label{1010}
\end{equation}
We sketch the calculation of photoconductivities 
for this rather general microwave field for $ a \ll R_c,\ell_B^2/\xi $, 
in App. \ref{apol}.
We obtain the result      
\begin{eqnarray} 
\sigma^{\rm photo} & = &   
\sum_{m \ge 1} \Theta\left[|m \omega_c-\omega|-\frac{\omega_c}{2}\right] (m \omega_c-\omega)^2 \; {\cal F}    \label{1020}    \nonumber \\
& & \times 
\left( {\cal R}_1 \;
\sigma^{\rm photo}(\Delta \omega \to m \omega_c-\omega, N)+
{\cal R}_2 \; 
\sigma^{\rm photo}(\Delta \omega \to m \omega_c-\omega, \nu) \right) 
\end{eqnarray}
where $ \sigma^{\rm photo}(\Delta \omega \to m \omega_c-\omega,\nu) $ 
is the placeholder for 
$ \sigma_{xx}^{\rm DP} $ 
$ \sigma_{yy}^{\rm DP} $, 
$ \sigma_{xx}^{\rm DF} $,
$ \sigma_{yy}^{\rm DF} $
calculated for the filling fraction $ \nu $ in the various regimes 
for  $ \Delta \omega/\omega_c \ll 1$. 
We carry out 
the replacement $ \Delta \omega \to m \omega_c-\omega $ 
in these conductivity 
expressions to get Eq. (\ref{1020}). 
We remark explicitly that this replacement 
should not be carried out in pure $ \omega $ terms in the 
photoconductivity expressions for 
$ V_N \gg \omega_c$ of section VI. 

The frequency dependent prefactor $ {\cal F} $ is given by 
 \begin{equation} 
{\cal F} =   
\left(
\frac{2 (\omega^2+\omega_c^2+2 \omega \omega_c\sin(2 \alpha) 
\sin(2 \beta))}{(\omega_c^2-\omega^2)^2}+ 
\mbox{po} \; \frac{(\omega_c^2-\omega^2)
\cos(2 \alpha)}{(\omega_c^2-\omega^2)^2}\right) \label{1040}\;. \\
\end{equation}  
The regime-dependent prefactors $ {\cal R}_1 $ and 
$ {\cal R}_2 $ are given by 
\begin{equation} 
{\cal R}_1 = \left\{
\begin{array}{ccc}
\frac{\omega_c}{\omega}-\left(\frac{\omega_c}{\omega}\right)^2 & 
\qquad \mbox{for} \qquad & T \ll V_N \ll \omega_c      \\ 
\frac{\omega_c}{\omega} & 
\qquad \mbox{for} \qquad & V_N\ll \omega_c, T \ll \mu      \\               0 & \mbox{for} & \omega_c \ll V_N
\end{array}    \quad ,    \right.  \label{1050}
\end{equation} 
and 
\begin{equation} 
{\cal R}_2 = \left\{
\begin{array}{ccc}
\left(\frac{\omega_c}{\omega} \right)^2  & 
\qquad \mbox{for} \qquad & T \ll V_N \ll \omega_c      \\ 
0  & 
\mbox{for} & V_N \ll \omega_c, T \ll  \mu      \\                         
\left( \frac{\omega_c}{\omega} \right)^2
 & \qquad \mbox{for} \qquad & \omega_c \ll V_N
\end{array}     \quad .   \right.  \label{1055}
\end{equation} 
The polarization-dependent factor $ {\rm po} $ is given by  
\begin{equation} 
\mbox{po}  = \left\{
\begin{array}{ccc}
1 & \qquad \mbox{for} \qquad & \sigma^{\rm photo}=
\sigma^{\rm DP}_{xx} \\
0 & \qquad \mbox{for} \qquad & 
\sigma^{\rm photo}=\sigma^{\rm DP}_{yy}, \sigma_{xx}^{\rm DF}\quad  
\mbox{or} \quad \sigma_{yy}^{\rm DF} \\
\end{array}        \right.     \,.        \label{1060} 
\end{equation} 

These conductivity formulas are only correct for $ a \ll R_c,\ell_B^2/\xi $.
The two different regime dependent prefactors 
$ {\cal R}_1 $ and $ {\cal R}_2 $ 
have their reason in the filling fraction dependence of the 
photoconductivity for $ T \ll V_N $.
We point out that the above replacement for the 
photoconductivities for $ V_N \gg \omega_c $ is simply 
a multiplication of the conductivity expression of section VI by 
the factor $ (\Delta \omega)^2 
(\omega_c/\omega)^2  {\cal F}$. The corresponding 
photoconductivity expressions are valid for all $ \omega>0 $. 

One can easily show 
that $ {\cal{F}} $ is a positive function which is 
important for the determination of the range of zero resistance states, and 
that $ {\cal{F}} $  
is only zero at the values $ \alpha=-\beta=\pm \pi/4 $ and 
$ \omega=\omega_c $. 
This corresponds to a circularly polarized  microwave field. 
From the calculation in App. E  one  further obtains
that for $ a \gg R_c $ or $ a \gg 
 \ell_B^2/\xi $ conductivity formulas 
similar to Eqs. (\ref{1020})-(\ref{1055}) hold, with a polarization factor 
$ \rm{po} $  depending on the specific impurity correlation function  
$ \tilde{W} $ which cannot be expressed in terms of the scattering times, as it was also the case for the photoconductivities in section IV. 

In Ref. \onlinecite{Dietel1} we calculated the microwave photoconductivity 
in the case of $ \omega \approx \omega_c $ for the linearly as well 
as circularly polarized microwave field and found no dependence of the 
photoconductivities on the polarization direction in the case of 
a linearly polarized microwave field. These results are in agreement with 
Eq. (\ref{1040}), where the last term in $ {\cal F} $ 
is approximately zero for 
$ \omega \approx \omega_c $. In contrast to this, for 
$ \omega $ not $ \approx \omega_c $, we find a polarization dependence only 
for   the displacement photoconductivity in $x$-direction. The 
 displacement photoconductivity in $y$-direction does not exhibit a 
polarization dependence.  

In the following, we compare the frequency terms ${\cal F} $ for $
\sigma^{\rm DP}_{xx} $ ($ \mbox{po}=1$) and $ \sigma^{\rm DF}_{xx} $
($ \mbox{po}=0$) with the corresponding terms of Vavilov {\it et al.}
\cite{Vavilov1} calculated for the case of conventional microwave
experiments without unidirectional periodic field. 
We find agreement \cite{Rem2} for  ${\cal F } $ (up to numbers)  
with the corresponding term of Ref. \onlinecite{Vavilov1}
for $ \sigma^{\rm DF}_{xx} $. The same holds true for 
$ \sigma_{xx}^{\rm DP} $ up to a sign 
difference in the second term in $ {\cal F} $ (\ref{1040}).
The prefactor of this term can be derived without explicit calculation 
as shown in App.~E.       
Finally, we note that Ref. \onlinecite{Vavilov1}  restricts the
calculation in the strong separated Landau level regime
to circularly polarized microwave fields. 

\section{Discussion}
We refer to section II for an overview and a discussion of the  
results of this paper. Beside this, we want to emphasis further 
the connection of this work and our earlier work Ref. \onlinecite{Dietel1}. 
One of the main theoretical consequences of 
Ref. \onlinecite{Dietel1} which was reconsidered in section III and 
summarized in the scaling relations (\ref{200}) was that  
the photoconductivities for $ a \ll R_c, \ell_B^2/\xi $ and 
$ V_N \ll \omega_c $ in the perpendicular direction 
to the modulation field are of the same order 
for the displacement as well as the distribution 
function mechanism.  
This is no longer the case for the striped state regime 
$ \xi \gg \ell_B^2/a $ where the photoconductivities 
of the distribution 
function mechanism are for both directions a factor 
$ (a/R_c)^2 (\tau_{\rm in} \tau^*_{\rm tr}/(\tau_{\rm s}^*)^2) $ (\ref{580}) 
larger than the photoconductivities of the displacement mechanism. 
We expect zero resistance states for both directions at certain 
frequency regions in the range    
$ \omega \approx \omega_c $. 
In the large overlapping Landau level regime the  
photoconductivity of the displacement mechanism in the 
perpendicular direction has no longer the oscillating property.  
Therefore we expect zero resistance states only parallel 
to the modulation direction. 

Summarizing, we obtain in the non-overlapping regime \cite{Dietel1} 
and in the overlapping Landau level regime parallel to the 
modulation direction accordance with the results 
of the photoconductivities without uniperiodic modulation potential. 
The concrete frequency 
dependence of the photoconductivities differ. For the system 
with modulation potential with weak disorder 
the photoconductivities can have large values at certain frequencies
having its origin in the singularities of the density of states 
at the band edge for the system without disorder.      
  
\begin{acknowledgements}
I would like to thank  F.~v.~Oppen and C.~Joas for useful discussions and the DFG-Schwerpunkt Quanten-Hall-Systeme for financial support. 
\end{acknowledgements}

\begin{appendix}

\section{The calculation of the photoconductivity $ 
{\boldmath \sigma_{yy}^{\rm DP}} $}
In this section we generalize the 
displacement photoconductivity formula 
in the y-direction for $ T \gg  V_N $ 
\cite{Dietel1} to a current formula 
without temperature restrictions.
The main difference between the calculation of the 
displacement photocurrent in $ x $ and in $ y $-direction 
consists on the fact that the 
many particle ground state without field can be represented 
as a single Slater determinant of eigenstates in the x-direction where in the 
y-direction we have a linear combination 
of Slater determinants built from the eigenstates. 
For a single Slater determinant of eigenstates we can use  
the standard Fermi's golden rule formula in order to calculate 
the current in $ x $-direction where one works with single Fermi factors 
\cite{Dietel1}.

As is well known the one particle eigenstates with a {\it dc} field in 
the x-direction can be represented 
in the Landau gauge by Hermite polynomials.  
As derived in \cite{Dietel1} the corresponding  one particle states 
with a {\it dc} field in the y-direction are  given by 
  \begin{equation}
  |n{\cal E}_l\rangle 
= \sqrt{2\pi \ell_B^2\over L_xL_y} \sum_k \exp\left\{i\int_0^k dk'{ {\cal E}_l
     - \epsilon^0_{nk'} \over eE_{\rm dc}} \right\}|nk\rangle. \label{a10}
\end{equation}
with the boundary condition $ {\cal E}_l = 
2\pi l eE_{\rm dc} \ell_B^2 /L_x $ and energies 
$ {\cal E}_{nl} = \hbar\omega_c (n + 1/2) + {\cal E}_l$.
The $ y $ expectation value of these states are given by 
$ \overline{y}({\cal E}_l) : =
\langle n {\cal E}_l | y | n {\cal E}_l\rangle =
- {\cal E}_{l} /eE_{\rm dc} $. 
As mentioned above the ground state wavefunction $ \Psi_0 $ 
is a Slater determinant 
of the one particle states $ |n k\rangle $. This can be 
represented as a linear combination of Slater determinants of 
$ | n {\cal E}_l \rangle $, i.e.
\begin{eqnarray}
\Psi_0 & = & \frac{1}{N!} \; S[|n_1 k_1\rangle,..,|n_N k_N\rangle] \nonumber \\
 & = & \sum\limits_{n_1 l_1<...<n_N l_N} a_{n_1 l_1;..;n_N l_N}
S[ |n_1 {\cal E}_{l_1}\rangle,..,|n_N {\cal E}_{l_N}\rangle ] \,. \label{a20}
\end{eqnarray}
Now we generalize the Fermi's golden rule current formula of 
Ref. \onlinecite{Dietel1} to the case that the ground state wavefunction 
consists on linear combinations of Slater determinants.
The current of such a state under microwave assisted 
impurity scattering is given by 
\begin{equation}
j_y^{\rm DP}= 
\sum\limits_{n_1 l_1<...<n_N l_N} |a_{n_1 l_1,..,n_N l_N}|^2 
j^{\rm DP}_{y;n_1 l_1,..,n_N l_N}                 \label{a30}
\end{equation} 
with the current components  
\begin{equation}
    j^{\rm DP}_{y;n_1 l_1;..;n_N l_N} 
     = {e\over L_xL_y} \sum_{nn'}\sum_{{\cal E}_l,{\cal E}_{l'}} |\gamma^\omega_{n{\cal E}_l\to n'{\cal E}_{l'}}|^2
     ({\overline y}({\cal E}_{l'}) - {\overline y}({\cal E}_l ))
     [ f^y(nl)_{n_1 l_1;..;n_N l_N}-
f^y(n'l')_{n_1 l_1;..;n_N l_N}  ] 
\delta({\cal E}_{nl} -{\cal E}_{n'l'} -\omega) \,.     \label{a35}
\end{equation}
Here $ f^y(n l)_{n_1 l_1;..;n_N l_N}=1 $ if 
$ (n,l) \in \{(n_1,l_1),..,(n_N,l_N)\} $ and zero otherwise. 
For the definition of the transition amplitude 
$ |\gamma^\omega_{n{\cal E}_l\to n'{\cal E}_{l'}}|^2 $ see 
Ref. \onlinecite{Dietel1}. 
The current formula (\ref{a35}) can be interpreted in the following way:
The total current is given by the sum of the 
probabilities of finding the electrons at the positions 
$ \overline{y}({\cal E}_{l_1}),.., \overline{y}({\cal E}_{l_N}) $ times the 
transition rate of these electrons through a surface of constant 
$ y $. We denote $  f^y_{nl} $ by 
\begin{equation} 
f^y_{n l}:= \frac{1}{N!} 
\sum\limits_{n_2 l_2<...<n_N l_N} 
|\langle \Psi_0 | S[ |n {\cal E}_{l}\rangle,|n_2 {\cal E}_{l_2}\rangle
,..,|n_N {\cal E}_{l_N}\rangle ]
\rangle|^2 \,.   \label{a40}
\end{equation}
By using the equations (\ref{a30})-(\ref{a40}) we obtain 
\begin{equation}
j_y^{\rm DP}=
{e\over L_xL_y} \sum_{nn'}\sum_{{\cal E}_l,{\cal E}_{l'}} 
|\gamma^\omega_{n{\cal E}_l\to n'{\cal E}_{l'}}|^2
     ({\overline y}({\cal E}_{l'}) - {\overline y}({\cal E}_l ))
     [ f^y_{nl}-
f^y_{n'l'}] 
\delta({\cal E}_{nl} -{\cal E}_{n'l'} -\omega)   \,.   \label{a45}
\end{equation}
In the following we show that $ f^y_{n l}=2 \pi \ell_B^2/(L_x L_y) 
\sum_k f(\epsilon^0_{nk})$. By using the definition 
$ | n {\cal E}_{l +L_x L_y /2 \pi \ell_B^2}\rangle 
:= | n {\cal E}_{l}\rangle  $ we  
will show first that $ f^y_{n l}=f^y_{n l'} $ for $ l,l' \in \mathbb Z $. 
This can be seen by using the one particle eigenstates for 
the y-direction (\ref{a10}) and the definition of the generalized 
Boltzmann factor (\ref{a40})  
\begin{eqnarray}
f^y_{nl} & = & \frac{1}{N!} 
\sum\limits_{n_2 l_2<...<n_N l_N} 
|\langle \Psi_0 e^{-i k_1({\cal E}_{l'}-{\cal E}_{l})/e E_{dc}} 
\cdots e^{-i k_N({\cal E}_{l'}-{\cal E}_{l})/e E_{dc}}
| S[ |n {\cal E}_{l}\rangle,|n_2 {\cal E}_{l_2}\rangle
,..,|n_N {\cal E}_{l_N}\rangle ]
\rangle|^2    \nonumber \\
& = & \frac{1}{N!} 
\sum\limits_{n_2 l_2<...<n_N l_N} 
|\langle \Psi_0 
| S[ |n {\cal E}_{l'}\rangle,|n_2 {\cal E}_{l_2+(l'-l)}\rangle
,..,|n_N {\cal E}_{l_N+(l'-l)}\rangle ]
\rangle|^2    \nonumber \\
&= & 
f^y_{nl'} .       \label{a50} 
\end{eqnarray}
By using that the number of states per Landau level is given by 
$ L_x L_y /(2 \pi \ell_B^2) $ we obtain (\ref{155})
for all $ l \in \mathbb{Z} $. 

\section{Some integrals of Laguerre polynomials \label{aint}}

In this paper we use a number of matrix elements  
of plane waves with respect to Laguerre polynomials in order to calculate 
the dark and photoconductivities. 
By the use of Ref. \onlinecite{MacDonald}
the  matrix elements are given by: 
\begin{eqnarray} 
&&  \int {d^2q\over (2\pi)^2} e^{-{q^2\ell_B^2\over 2}} 
[L^0_{n} ({q^2\ell_B^2\over 2})]^2(q_y \ell_B)^{2 r}   =  
{1\over 2 \pi\ell_B^2} (\delta_{r,0} + \delta_{r,1} 2 n) 
         \,,  \nonumber \\
 & &  \int {d^2q\over (2\pi)^2} e^{-{q^2\ell_B^2\over 2}} [ L^m_{n+ 1}
({q^2\ell_B^2\over 2})-L^m_n({q^2\ell_B^2\over 2})]^2 
\left(\frac{(q \ell_B)^2}{2 n}\right)^m (q_y \ell_B)^{2 r} 
=  
 {1\over 2 \pi\ell_B^2} 
(\delta_{r,0} 2 + \delta_{r,1}  6n ) \,,
\nonumber \\
&&\int {d^2q\over (2\pi)^2} e^{-{q^2\ell_B^2\over 2}} 
[ L^{m-1}_{n+ 1} ({q^2\ell_B^2\over 2})-L^{m-1}_n({q^2\ell_B^2\over 2})] 
[ L^{m+1}_{n- 1} ({q^2\ell_B^2\over 2})-L^{m+1}_n({q^2\ell_B^2\over 2})] 
\nonumber \\
&& \times 
\left(\frac{(q \ell_B)^2}{2 n}\right)^{m-1}
\left(\frac{((q_x \pm iq_y) \ell_B)^2}{2 n}\right) (q_y \ell_B)^{2 r}
 =  
{1\over 2 \pi\ell_B^2} (\delta_{r,0} \,0- \delta_{r,1} 3n )  \label{a500}
\end{eqnarray}
 for  $ r=0$ or $ r=1$, respectively. The results above are correct 
in the leading $ n $ order for $ n \to \infty$.

 \section{The asymptotic expansions of the Laguerre polynomials}
In this section we review some of the known asymptotic expansions of 
the Laguerre 
polynomials. A full description can be found in \cite{Erdelyi}.
In the following we give expansions for large $ n $ of 
$ e^{-x/2} L_n^{\alpha}(x) $ for $ x $ near $ 0 $, in the oscillatory 
region, and in the monotonic region. \\
    
The leading order term of an expansion 
which is valid for {\it x  near zero}, i.e. 
$ 0\le x<n^{1/3} $, is given by   
\begin{equation}
e^{-x/2} L_n^{\alpha}(x)= 
\frac{\Gamma(n+\alpha+1)}{\left(\frac{\nu}{4}\right)^{\frac{\alpha}{2}}n!
\, x^{\frac{\alpha}{2}}}\,
J_{\alpha}[(\nu x)^{\frac{1}{2}}]+O\left(n^{\frac{\alpha}{2}-\frac{1}{4}}\right)   \label{a600}
\end{equation}
where $ \nu=4 n +2 \alpha +2 $.
One can specialize this expansion in the range $ 0<\epsilon < x<n^{1/3} $
with $ \epsilon >0 $ to Fej\'{e}r's formula 
\begin{equation}
e^{-x/2} L_n^{\alpha}(x)
=\frac{n^{\frac{\alpha}{2}-\frac{1}{4}}}
{\pi^{\frac{1}{2}}x^{\frac{\alpha}{2}+\frac{1}{4}}}
\cos\left[\,2(nx)^{\frac{1}{2}}-\alpha \frac{\pi}{2}-
\frac{\pi}{4}\right]+O\left(n^{\frac{\alpha}{2}-\frac{1}{4}}\right)\,.
\label{a610}
\end{equation} 
With the help of 
\begin{equation} 
x= \nu \cos^2(\theta_\nu)\,,  \qquad 0<\theta_\nu<\pi/2 
\,,  \qquad 4\Theta_\nu(x)=\nu(2 \theta_\mu-\sin(2 \theta_\nu))+\pi     \label{a630}
\end{equation} 
the Laguerre polynomials in the {\it oscillatory region}  
$ 0 <x < \nu $ are given by   
\begin{equation} 
e^{-x/2} L^\alpha_n(x)=2 (-1)^n 
\frac{1}{(2 \cos(\theta_\nu))^{\alpha}}
\frac{1}{(\pi \nu \sin( 2 \theta_\nu))^{\frac{1}{2}}}
\left(\sin(\Theta_\nu)+O\left(\frac{1}{n}\right)\right) \,.                              \label{a640}
\end{equation}

One can find similar asymptotic expansions of $ e^{-x/2} L^\alpha_n(x) $ 
at the transition point $ x \approx \nu $ and in the monotonic region 
$ x > \nu $ (see e.g. \cite{Erdelyi}). Because both regions give for 
$ n \to \infty $ vanishing 
contributions to the calculated physical quantities in this paper 
we shorten here the discussion of the asymptotics in 
these regions.    
In the monotonic region $ x> \nu $ the  
function $e^{-x/2} L^\alpha_n(x) $ behaves asymptotically as  
$ e^{-\Theta_\nu'} $ 
with $ \Theta_\nu'=\nu(\sinh(2\theta_\nu')-2 \theta_\nu') $ and 
$ x=\nu \cosh^2(\theta_\nu') $, $ \theta_\nu'>0 $ (see \cite{Erdelyi}) .    
Furthermore, one can show that both  asymptotical expansions, i.e.  
in the oscillatory region as well as in the monotonic region, are  
valid at the transition point except in the  range $ |x-\nu| < \nu^{1/3} $.

\section{The calculation  
of plane waves matrix elements with respect to WKB eigenstates}
In this section, we calculate the matrix elements 
$ \left<nk|e^{i \vec{q} \vec{r}}|n'k'\right> $.
Before going on we will first quote for comparison 
the well known result of this matrix element for $ \tilde{V}=0 $ 
by using the exact Hermite polynomial eigenfunctions. With 
\begin{equation}
A_{n,n'}(\vec{r})=\sqrt{\frac{n'!}{n!}}\left(\frac{x+i y}{\sqrt{2} \ell_B} 
\right)^{n-n'} L_{n'}^{n-n'}\left(\frac{r^2}{2\ell_B^2}\right)   \label{a1000}  
\end{equation}
for $ n \geq n' $ and $ A_{n,n'}(\vec{r})=(-1)^{n-n'}
A_{n',n}^*(\vec{r})$ for 
$ n' \geq n $
we obtain
\begin{align}
& \left<nk|e^{i \vec{q} \vec{r}}|n'k'\right>  =  \nonumber  \\ 
& \delta(k'-k+q_y) \, e^{i q_x(k+k')\ell_B^2/2}\,  i^{n-n'} 
e^{-q^2 \ell_B^2/4} A_{n,n'}(\vec{q}\ell_B^2) \,.      \label{a1005}  
\end{align} 

Now we come to the calculation of this matrix element in the 
WKB approximation: 

For simplification, we denote $ \Theta^V_{n} $ (\ref{2060}), 
$  C^{V}_{nk} $ (\ref{2070}) as 
a function of $ x $ by 
$ \Theta^V_{n}(x) $, $  C^{V}_{nk}(x) $;  as a function of 
$x'=x/\sqrt{n+1/2}\ell_B $ by $ \tilde{\Theta}^V_{n}(x') $, 
$  \tilde{C}^{V}_{nk}(x') $; and as functions of 
$ \varphi_x=\arcsin(x/\sqrt{2(n+1/2)} \ell_B) $ by 
$ \tilde{\tilde{\Theta}}^V_{n}(\varphi_x) $, 
$  \tilde{\tilde{C}}^{V}_{nk}(\varphi_x) $. 
Here $ \arcsin(x) \in [-\pi/2,\pi/2] $. Furthermore, we extend 
the range of definition of $ \tilde{\tilde{\Theta}}^V_{n}(\varphi) $ and  
$  \tilde{\tilde{C}}^{V}_{nk}(\varphi) $ to $ [-\pi,\pi] $ by 
$ \tilde{\tilde{\Theta}}^V_{n}(\varphi):=
\tilde{\tilde{\Theta}}^V_{n}(\arcsin(\sin(\varphi))) $ and  
$ \tilde{\tilde{C}}^{V}_{nk}(\varphi):=
\tilde{\tilde{C}}^{V}_{nk}(\arcsin(\sin(\varphi))) $ for all 
$ \varphi \in [-\pi,\pi] $.

First,  by using (\ref{2060}) we get for $n \to \infty$ 
\begin{align} 
& \nts{2}\left<nk|e^{i \vec{q} \vec{r}}|n'k'\right>=
\frac{\ell_B^2}{N^2_{\Psi}} \delta(k'-k+q_y) 
e^{iq_x(k+k')\ell_B^2/2} \label{a1010}\\
& 
\times \int\limits_{-1+|\tilde{q}_y|/2}^{1-|\tilde{q}_y|/2} dx'
e^{i (n+1/2)\,\tilde{q}_x x'} 
\left(\frac{1}{1-(x'-|\tilde{q}_y|/2)^2} \right)^{\frac{1}{4}}
\left(\frac{1}{1-(x'+|\tilde{q}_y|/2)^2 }\right)^{\frac{1}{4}}  
\nonumber \\
& \times 
\sin\left(\tilde{\Theta}^V_{n}(x'-
\tilde{q}_y/2)+\tilde{C}^{V}_{nk}(x'-\tilde{q}_y/2)+
\frac{\pi}{4}\right) 
\sin\left( \tilde{\Theta}^V_{n'}(x'+\tilde{q}_y/2)+
\tilde{C}^{V}_{n'k'}(x'+\tilde{q}_y/2)+\frac{\pi}{4}\right)    \nonumber 
\end{align} 
with $ \tilde{q}_x=q_x \ell_B/ \sqrt{2(n+1/2)}  $ and 
$ \tilde{q}_y=q_y \ell_B /\sqrt{2(n+1/2)}    $. 

In the following, we calculate first the matrix element (\ref{a1005}) 
for $ q \ell_b $ {\it near zero}, i.e. $q^2 \ell^2_b \ll n^{1/3}$. 
Similar as in (\ref{2090}) we obtain for fixed $ \Delta n $, 
$ (\Delta x/\ell_B)^2 \ll n^{1/3}  $ and   
$ n  \to \infty $ 
 \begin{equation}
 \Theta^V_{n+\Delta n}(x-\Delta x)\approx \Theta^V_{n}(x)+
 2(n+1/2)\sqrt{1-\frac{x^2}{2 (n+1/2)\ell_B^2}}
\;  \frac{\Delta x}{\sqrt{2(n+1/2)}\, \ell_B} +
 \arccos\left(\frac{x}{\sqrt{2(n+1/2)}\, \ell_B} \right) 
\Delta n  \,.  \label{a1020}
 \end{equation}
Now we use 
this expansion in (\ref{a1010}). 
With the assumption $ q_x \ell_B/\sqrt{2 (n+1/2)} \ll 1 $ we obtain  
\begin{align} 
&   \left<nk|e^{i \vec{q} \vec{r}}|n'k'\right> = 
\delta(k'-k+q_y) e^{iq_x(k+k')\ell_B^2/2} \frac{1}{\pi}
\int\limits_{-\pi/2}^{\pi/2}\; d\varphi_x
 e^{i (q_x \ell_B) \sin(\varphi_x) \sqrt{2(n+1/2)}}  \nonumber \\
& \times \cos\left[2 (n+1/2) \cos(\varphi_x) 
\frac{(k-k')\ell_B}{\sqrt{2 (n+1/2)}}+ (n-n')
\left(\frac{\pi}{2} -\varphi_x\right)
+\tilde{\tilde{C}}^V_{nk}(\varphi_x)-
\tilde{\tilde{C}}^V_{nk'}(\varphi_x)
\right]     \,.                 \label{a1030}
\end{align}     
When carrying out the integral in (\ref{a1030}) 
for $ C^V_{nk}=C^V_{nk'}=0 $ we obtain the exact result 
(\ref{a1005}) when inserting the asymptotics (\ref{a600})
for the Laguerre polynomial.

Next, we calculate $ \left<nk|e^{i \vec{q} \vec{r}}|n'k'\right> $ 
from (\ref{a1030}) for $ C^V_{nk}\not=0 $. 
It is difficult to calculate the integral in this case 
without approximation. But for $ \sqrt{(k-k')^2 +q_x^2 }\ell_B>0 $ one can 
get the leading term for $ n \to \infty $ with the help of a 
saddle point approximation. For doing this we first split the cosine 
in its two exponential terms. Then by carrying out a 
saddle point approximation  we get by the help of  
the definitions  $ \cos(\varphi_0):=(k-k')/\sqrt{q_x^2 +q_y^2}$ 
and $ \sin(\varphi_0):=-q_x/\sqrt{q_x^2 +q_y^2 }$ for 
$ k-k' \ge 0 $ and no restrictions on $ n $, $ n' $  
\begin{align}
&  \left<nk|e^{i \vec{q} \vec{r}}|n'k'\right> = 
\delta(k'-k+q_y) e^{iq_x(k+k')\ell_B^2/2}
\frac{1}{\sqrt{2\pi}} 
\left(\frac{1}{2(n+1/2)q \ell_B}\right)^{\frac{1}{4}} \label{a1040}\\
& \times e^{i(n-n') \varphi_0}   
\sum_{\sigma \in \{\pm\}} 
\exp \left[i \sigma \left(\sqrt{2(n+1/2)}q\ell_B+\left((n-n')
-\frac{1}{2}\right)\frac{\pi}{2}
+\tilde{\tilde{C}}^V_{nk}(-\sigma \varphi_0) -
\tilde{\tilde{C}}^V_{nk'}(-\sigma \varphi_0) \right) \right] \,.\nonumber     
\end{align}  
By comparing (\ref{a1040}) for $ C^V_{nk}=C^V_{nk'}=0 $ 
with (\ref{a1005}) when inserting for the Laguerre polynomial  
the asymptotics (\ref{a610}) we get correspondence of both formulas. \\

Next, we calculate $  \left<nk|e^{i \vec{q} \vec{r}}|n'k'\right> $ 
in the case where $ q^2 \ell_B^2/8 (n+1/2) <1 $ is larger than zero 
for $ n \to \infty$.
This will be done, by carrying out a saddle point approximation for 
the integral in (\ref{a1010}) when the product of the 
two sinus terms are split into its
four exponential parts. Then one gets after some algebraic manipulation 
 by the use of 
(\ref{a1020}) that two of these  exponents have stationary points 
only in the case 
when $ q^2 \ell_B^2/8 (n+1/2) <1 $ ($ n \to \infty $). 
According to (\ref{a640}) this corresponds 
to the {\it oscillatory region} of the Laguerre polynomials with argument 
$  q^2 \ell_B^2/2$. 
Which of the two exponential terms exhibit  the stationary points 
depends on the relation of $ \tilde{q}^2 $ to $ 2 \tilde{q}_y $.  
These two stationary points of the exponents in 
(\ref{a1010}) are given by 
$x'_s = \pm (\tilde{q}_y/\tilde{q}) \sqrt{1-\tilde{q}^2/4} $.
Additionally to the angle  $ \varphi_0 $ defined above (\ref{a1040}) it is 
useful to define the angle $ \varphi_{\tilde{q}}  $ by 
$ \sin(\varphi_{\tilde{q}}):= \sqrt{1-\tilde{q}^2/4} $ and 
$ \cos(\varphi_{\tilde{q}}):= \tilde{q}/2 $.
After some tedious calculation we get in the leading order for  
$ n \to \infty $ and  $ k-k'>0 $
 \begin{align}
 &  \left<nk|e^{i \vec{q} \vec{r}}|n'k'\right> = 
 \delta(k'-k+q_y) \, e^{iq_x(k+k')\ell_B^2/2}
\,2 \, (-1)^{n} \, 
 e^{i(n-n') \varphi_0} 
\left(\frac{1}{\pi \nu \sin\left( 2 \, 
\theta_\nu \nts{2}\left(\frac{q^2 \ell_B^2}{2}\right)\right)}
\right)^{\frac{1}{2}}    \,.   \label{a1050}\\
 & 
\times  
 \sum_{\sigma \in \{\pm\}} 
 \exp \bigg[i \sigma \; \Theta_\nu \nts{2}\left(\frac{q^2 \ell_B^2}{2}\right) 
+i c_{\sigma,k}\; 
\tilde{\tilde{C}}^V_{nk}\left(-
  \left(\varphi_0+\frac{\pi}{2}\right)-\sigma \varphi_{\tilde{q}}\right)
+i c_{\sigma,k'}\; \tilde{\tilde{C}}^V_{nk'}\left(+
 \left(\varphi_0+\frac{\pi}{2}\right)-\sigma \varphi_{\tilde{q}}\right)
\bigg)\bigg] \nonumber     
 \end{align} 
with $ \nu=2(n+n')+2 $  where $ \theta_\nu $ and  $ \Theta_\nu $ are  defined in (\ref{a630}).
The matrix $ c_{\sigma,k} $ is defined by 
$ c_{-,k'}= c_{+,k} $ and $ c_{-,k}= c_{+,k'}$. Furthermore, we have 
$ c_{+,k}=c_{+,k'}= q_x/|q_x| $ for 
$ |\varphi_0|\ge |\varphi_q| $. $ c_{+,k}=-1 $, $ c_{+,k'}=1 $ for 
$ |\varphi_q|\ge |\varphi_0| $.

Finally, we compare the WKB formula (\ref{a1050}) for 
$ C^V_{nk}=C^V_{nk'}=0 $ 
with the exact formula (\ref{a1005}) when inserting in the 
Laguerre polynomial the asymptotics (\ref{a640}). 
We get correspondence of both formulas.

\section{The calculation 
of the photoconductivities in the case of arbitrary frequency 
and polarization of the microwave field \label{apol}}

Up to now, we restricted the determination of the photoconductivities to the 
case where the system is irradiated by a microwave field 
polarized in $x$-direction of frequency 
$ \omega \approx \omega_c $.  
The knowledge of the polarization dependency under microwave irradiation
could be useful when determining the physical mechanism 
of the microwave oscillations \cite{Smet1}.  
Therefore we calculate in this section 
the various photoconductivities in the case 
that the irradiation has the rather 
general form 
$ \vec{E}=\mbox{Re}[\vec{\cal{E}} e^{-i \omega t}]$.
Here $ \vec{\cal{E}} $ is 
a two dimensional vector with complex values. This type of electric fields 
can be represented by the following potential  
\begin{equation}
\phi(\alpha,\beta) = 
- eE \left(x \cos(\alpha) \cos(\omega t-\beta ) + 
y \sin(\alpha) \cos(\omega t+\beta)\right) =
\phi_{+}(\alpha,\beta)e^{-i\omega t} +\phi_{-}(\alpha,\beta)
e^{i\omega t}  \,.  \label{a200}
\end{equation}
Because we calculate {\it dc} response values we fixed   
the freedom of the
$ t=t_0 $ starting point of the plane wave in (\ref{a200}) such that the 
phase factor of the x-component of the electric field is minus the 
phase factor  of the y-component.   
$ \phi_{+}(\alpha,\beta) $ is then given by 
\begin{equation}
\phi_{+}(\alpha,\beta)=-{eE\over 2} 
\left(x \cos(\alpha) e^{i \beta}+y \sin(\alpha)
e^{-i \beta}\right) \label{a210}
\end{equation} 
and $ \phi_{-}(\alpha,\beta)=\phi^*_{+}(\alpha,\beta) $.

In the following, we give only a short outline of the calculation of 
the photoconductivities in the case $ V_N \ll \omega_c $, $ a \ll R_c $ 
and $ T \gg V_N $ which 
was the regime discussed in \cite{Dietel1}.  
By using the results for the transition matrix elements for the x-polarized  
and y-polarized  microwave field calculated in \cite{Dietel1} we obtain 
 \begin{eqnarray}
     |\langle n\pm m  k' | T_\pm(\alpha,\beta) |
  n k \rangle|^2
     &=& \left({eER_c\over 4}\right)^2 
       \int {d^2q\over (2\pi)^2} \,\delta_{q_y,k'-k} e^{-{q^2\ell_B^2\over 2}} 
  \langle U(\vec{q})U(-\vec{q})\rangle \label{a220} \\
& & 
 \times \bigg| \left(\cos(\alpha) e^{i \beta} +i\sin(\alpha) e^{-i\beta}
\right) M_1(\vec{q}\,\ell_B)+\left(\cos(\alpha) e^{i \beta} 
-i \sin(\alpha) e^{-i\beta}
\right) M_2(\vec{q}\,\ell_B)\bigg|^2 \nonumber 
  \end{eqnarray} 
with
\begin{eqnarray}
M_1(\vec{q}\,\ell_B) & = &  \frac{\omega_c}{\omega(\omega_c-\omega)} 
\left(\frac{(-q_y+iq_x)\ell_B}{\sqrt{2 n}}\right)^{|m|-1} 
  [ L^{|m|-1}_{n+1} ({q^2\ell_B^2\over 2})
         -L^{|m|-1}_{n}({q^2\ell_B^2\over 2})] \,, \label{a225} \\
M_2(\vec{q}\,\ell_B) & = & \frac{\omega_c}{\omega(\omega_c+\omega)} 
 \left(\frac{(-q_y+iq_x)\ell_B}{\sqrt{2 n}}\right)^{|m|+1} 
  [ L^{|m|+1}_{n-1} ({q^2\ell_B^2\over 2})
        -L^{|m|+1}_{n}({q^2\ell_B^2\over 2})] \,. \label{a230}
\end{eqnarray} 
Here $ T_\pm(\alpha,\beta) $ is the second order transfer matrix 
\cite{Dietel1} containing the impurity potential 
and $ \phi_{\pm}(\alpha,\beta) $. The calculation of the matrix 
elements was carried out by the insertion of intermediate states 
in between the microwave operator and the impurity operator 
in $ T_\pm(\alpha,\beta) $ 
\cite{Dietel1}.  
$ M_1 $ in (\ref{a225}) was 
already considered before when we have calculated the photoconductivities 
for $ \omega \approx \omega_c $. This term comes from intermediate states 
of Landau level index $ n+1 $ and $ n+m-1 $. $ M_2 $ in 
(\ref{a230})  comes from 
matrix elements with intermediate states of Landau level 
index $ n-1 $ and $ n+m+1 $. 
It then immediately clear that the prefactor of $ M_2 $ can be deduced from 
the prefactor of $ M_1 $ by the substitution 
$ \omega_c \rightarrow -\omega_c $.   
  
By using the matrix elements in appendix \ref{aint} we obtain 
for delta correlated impurities 
\begin{equation} 
\sum_k\;  |\langle n\pm m  k' | T_\pm(\alpha,\beta) |  n k \rangle|^2 =  
  \left({eER_c\over 4}\right)^2 {1\over 2\pi \nu_0 \tau_{\rm tr}} 
\frac{1}{ \pi \ell_B^2} 
\left(\frac{\omega_c}{\omega}\right)^2 
\frac{2(\omega^2+\omega_c^2+2 \omega \omega_c\sin(2 \alpha) 
\sin(2 \beta))}{(\omega_c^2-\omega^2)^2}  \label{a270}
\end{equation}
and 
\begin{eqnarray}
& &  \sum_k\; |\langle n\pm m  k' | T_\pm(\alpha,\beta) |  n k \rangle|^2 
 ((k-k')\ell_B)^2  \nonumber \\
& & 
 = 
\left({eER_c\over 4}\right)^2 {\tau_{\rm s}\over 2\pi \nu_0 \tau^2_{\rm tr}} 
\frac{6n}{2 \pi \ell_B^2} \left(\frac{\omega_c}{\omega}\right)^2 
 \left(
\frac{2 (\omega_c^2+\omega^2+2 \omega_c \omega \sin(2 \alpha) 
\sin(2 \beta))}{(\omega_c^2-\omega^2)^2}+\frac{(\omega_c^2-\omega^2)
\cos(2 \alpha)}{(\omega_c^2-\omega^2)^2}\right) \,.  \label{a280}
\end{eqnarray}
By using the identity 
$   (x/n) e^{-x/2} L^m_n(x)=- e^{-x/2} L^{m-2}_{n+1}(x) $ 
which is valid for $ n \to \infty $ (\ref{a640}) and further by taking  
into account the considerations in \cite{Dietel1}  
we obtain that (\ref{a270}) and (\ref{a280}) are also valid for non delta 
correlated impurities.

The last term in (\ref{a280}) stems from the term proportional to 
$ M_1 M_2 $ in the integrand in (\ref{a220}) resulting in the 
polarization dependency of $ \sigma_{xx}^{\rm DP}$. This 
interference term $  M_1 M_2 $ does not exist in 
$ \sum_k\; |\langle n\pm m  k' | T_\pm(\alpha,\beta) |  n k \rangle|^2$ 
(\ref{a270}).
By using the results above and further that 
\begin{equation} 
\sum_n (f_n-f_{n+m}) = m             \;.       \label{a290}
\end{equation} 
we obtain for the photoconductivities (\ref{1020}). The generalization 
of the calculation to the other regimes in (\ref{1020}) is straight
forward.

\end{appendix}


\begin{thebibliography} {50}
\bibitem{Rem1} 
We note that the numerical prefactors of $ \sigma_{xx}^{\rm DP}$, 
and $ \sigma_{yy}^{\rm DF} $ differ from the 
 corresponding expressions in \cite{Dietel1} due to miscalculations.
\bibitem{Rem2}
In order to show that the frequency dependent prefactor $ {\cal F} $ 
is in accordance with the results in Ref. \onlinecite{Vavilov1} 
for the system without modulation we use that $ \sin(2
\alpha) \sin(2 \beta) $ in the first numerator of Eq.(\ref{1040}) is
equal to $ -2 \mbox{Im}[{\cal E}^*_x {\cal E}_y]/ |{\cal E}|^2 $ and 
$ \cos(2 \alpha) =  (|{\cal E}_x|^2-|{\cal E}_y|^2)/ |{\cal E}|^2 $ 
where
$ \vec{\cal{E}} $ is the complex two-dimensional vector which is
defined by $ \vec{E}(\alpha,\beta)=\mbox{Re}[\vec{\cal{E}}e^{-i \omega
  t}]$ .  When inserting the definitions of $ {\cal E}_x $ and $ {\cal
  E}_y $ of Ref. \onlinecite{Vavilov1} in Eq.
(\ref{1040}) we find agreement.
\end{thebibliography}
\end{document}